\documentclass[a4paper,11pt]{article}
\usepackage{jheppub}
\usepackage{amsmath}
\input epsf
\usepackage{epsfig}
\usepackage{amssymb}
\usepackage{graphics}
\usepackage[active]{srcltx}
\usepackage{cancel}

\usepackage{setspace}
\usepackage{array,multirow,bigdelim,arydshln}
\usepackage{shuffle}

\usepackage{bm}

\usepackage{listings}
\usepackage{xcolor}
\usepackage{tcolorbox}
\tcbuselibrary{listingsutf8}

\newcommand{\be}{\begin{equation}}
	\newcommand{\ee}{\end{equation}}
\newcommand{\bea}{\begin{eqnarray}}
	\newcommand{\eea}{\end{eqnarray}}
\newcommand{\bean}{\begin{eqnarray*}}
	\newcommand{\eean}{\end{eqnarray*}}
\newcommand{\nn}{\nonumber \\}

\def\W #1{\widetilde{#1}}
\def\WH #1{\widehat{#1}}

\def\d{\partial}

\def\eps{\epsilon}
\def\a{{\alpha}}

\def\eref#1{(\ref{#1})}

\def\Label#1{\label{#1}%
	\smash{\hbox to0pt{\raise1ex\hbox{\tiny[#1]}\hss}}}

\def\co{\,,}

\title{General One-loop Generating Function by IBP relations}
\author{Chang Hu$^{abcde}$\footnote{The non-standard author order is for young researchers to gain recognition from their institutions for this work by the outdated practice in China. }, Bo Feng$^{fg}$\footnote{Emails:  fengbo@csrc.ac.cn, isiahalbert@126.com, isaiahyshen@gmail.com, yaobozhang@nxu.edu.cn},  Jiyuan Shen $^{deh}$, Yaobo Zhang$^{i}$ \\~\\
	{$^a$\small College of Physics Science and Technology, Hebei University, Baoding 071002, China\\
	$^b$\small Hebei Key Laboratory of High-precision Computation and Application of Quantum Field Theory, Baoding, 071002, China\\
	$^c$\small Hebei Research Center of the Basic Discipline for Computational Physics, Baoding, 071002, China\\
	$^d$\small School of Fundamental Physics and Mathematical Sciences, Hangzhou Institute for Advanced Study, UCAS,
	Hangzhou 310024, China\\
	$^e$\small University of Chinese Academy of Sciences, Beijing 100049, China \\
	$^f$\small Beijing Computational Science Research Center, Beijing 100193, China \\
	$^g$\small Peng Huanwu Center for Fundamental Theory, Hefei, Anhui, 230026, China \\
	$^h$\small Institute of Theoretical Physics, Chinese Academy of Sciences, Beijing 100190, China \\
	$^i$\small School of Physics, Ningxia University, Yinchuan, 750002, China
	}
}
\date{\today}
\abstract{
	In this paper, we have studied the general generating function of reduction for one-loop integrals with arbitrary
	tensor structure in the numerator and arbitrary power distribution of propagators in the denominator. Using  IBP relations, we have established
	the partial differential equations for these generating functions and solved them analytically with compact form. These results provide 
	useful information for the analytical structure of reduction coefficients and guidance for applying the generating function method to reductions of higher loop integrals.
	
}


\newpage
\begin{document}

\maketitle
\tableofcontents

\section{Motivation}

As a connection between the theoretical framework and experimental data, the scattering amplitude is a central concept of quantum field theory,
thus its efficient computation is extremely important. For tree-level amplitudes, the on-shell recursion relation \cite{Britto:2004ap, Britto:2005fq} has marked a significant advancement in both new efficient computational tools and a fresh on-shell perspective for studying quantum field theory\footnote{For the introduction of these
	new developments, please see following books \cite{Henn-book, Elvang-book, Badger-book}}. Compared to tree-level amplitudes,
the computation of loop amplitudes is much more difficult. Many efforts have been devoted to this subject (see the books \cite{Smirnov:2004ym, Smirnov:2012gma}).

A fundamental idea of computing loop amplitudes is the reduction of loop integrals, which states that any
loop integral can be written as the linear combination of some basic integrals (called the master
integrals) with coefficients (called the reduction coefficients) as the rational function of external momenta, polarization vectors, masses and spacetime dimension. Based on this observation, loop integrals have been divided into two relatively independent parts, i.e., the integration of master integrals and the abstraction of reduction coefficients.
Progress in either part will increase the efficiency of computing loop amplitudes. A nice introduction to
recent achievements can be found in the book  \cite{Weinzierl:2022eaz}. For one-loop integrals, the basis is well known around 2000
(see the reference \cite{Ellis:2007qk} and the code implement \cite{Carrazza:2016gav} ), the one-loop revolution comes,
in fact,  from the breakthrough in the reduction methods, among which the OPP-method \cite{Ossola:2006us} and the unitarity cut method \cite{Bern:1994zx, Bern:1994cg, Britto:2004nc} are two main representative methods.

In this paper, we will focus on the analytic abstraction of reduction coefficients. When we talk about the reduction, there
are, in fact, two different categories\footnote{Although we have roughly divided reductions into two categories, they are, in fact,
	mutual influence. For example, OPP method learns a lot from the unitarity cut method and the unitarity cut method can also be applied to the reduction at the integrand level, where the product of on-shell tree-level amplitudes has been used to provide the integrand for the reduction as shown in \cite{Bern:1994zx, Bern:1994cg, Ossola:2006us}.	 
}. The first category is the algebraic decomposition of loop integrands, which we will call
the "integrand level reduction" method. The key idea is to write a complicated rational function into several simpler pieces.
For one-loop integrals,  the well-known OPP-method \cite{Ossola:2006us} is a representative method in this category. 
After realizing the splitting of multivariable rational functions can be systematically solved using the computational algebraic geometry \cite{Mastrolia:2011pr, Badger:2012dp, Zhang:2012ce}, the integrand level reduction for higher loop integrals is solved in principle. Although it is highly systematic, one drawback is the presence of large
amounts of spurious terms, which contribute nothing physically but are necessary for calculation in this framework.
For one-loop integrals, this drawback is not so severe. However, for two and higher loops, it becomes a severe problem. 
To avoid
this inefficiency, the second category of reduction method becomes more attractive. Since only reduction coefficients of the integrand basis
with nonzero contributions after integration 
are computed, we name this category the "integral level reduction" method.

Historically, the first realization of reduction is done at the integral level, which was proposed by Passarino and
Veltman in \cite{Passarino:1978jh} and is called the PV-reduction method. Since then, other alternative integral level reduction
methods have been proposed. Some widely mentioned methods include  the
Integration-by-Part (IBP) method \cite{Chetyrkin:1981qh,Tkachov:1981wb,Laporta:2000dsw,vonManteuffel:2012np,vonManteuffel:2014ixa,Maierhofer:2017gsa,Smirnov:2019qkx},
the unitarity cut method \cite{Bern:1994zx,Bern:1994cg,Britto:2004nc,Britto:2005ha,Britto:2006sj,Anastasiou:2006jv,Anastasiou:2006gt,Britto:2006fc,
	Britto:2007tt,Britto:2010um},
the intersection theory method \cite{Mastrolia:2018uzb, Frellesvig:2019uqt,Mizera:2019vvs, Frellesvig:2020qot,Caron-Huot:2021xqj, Caron-Huot:2021iev}. It is worth
mentioning that different reduction methods utilize different properties of loop integrals. The PV-reduction method has
used the tensor structure of loop integrals. The IBP method has used the translation and scaling properties of loop integrals under
dimensional regularization. The unitarity cut method has used the analytic property (such as the imaginary part and the leading singularity etc) of loop integrals. The intersection theory method
has used the deep mathematical insight of twist cohomology and dual twist cohomology structure of loop integrals
\cite{Inter-book, Cho:2016wxs, Mizera:2017rqa}. It is interesting to see if we could explore other properties of loop integrals
or combine these different methods to get new reduction methods.

No matter which reduction method one uses, a notable common feature in the reduction procedure is the appearance of an iterative structure.
For example, by IBP relation, one can write the integral of higher rank tensor to the linear combination of integrals of
lower rank tensors. Wisely utilizing
these iterative structures without repeatedly computing them is an important step in enhancing the efficiency of reduction\footnote{About the related discussions, see some recent papers  \cite{Chen:2022jux, Chen:2022lue, Zhang:2023jzv}}. When encountering an iterative structure, a very useful approach is to consider the corresponding
{\bf generating function}. In many examples, it is much easier to find the generating function rather than to find the expansion coefficients
of each order.  For example, in \cite{Ablinger:2014yaa},  the generating function for an operator insertion on an $l$-leg vertex
was introduced, which facilitates the computation of operator matrix elements of higher loop contributions. Similarly in \cite{Kosower:2018obg}
generating functions for a specific tensor structure of some two-loop integrals have been introduced and solved, which
provide explicit reductions of that tensor structure with arbitrary power.

One of our ultimate goals is to solve the reduction of general loop integrals, i.e., integrals with the general polynomial of loop momenta in the numerator
and the general power distribution of propagators in the denominator. 
 A general numerator implies a general tensor structure, which is the
starting point of the PV-reduction method. Recently, to simplify the manipulation of tensor
structure,   an auxiliary vector $R$
has been introduced in \cite{Feng:2021enk,Hu:2021nia,Feng:2022uqp,Feng:2022iuc,Feng:2022rfz,Li:2022cbx,Chen:2022jux, Chen:2022lue, Zhang:2023jzv}. Taking  advantage of this method, the generating function for
tensor reduction of one-loop integrals has been constructed and solved in \cite{Feng:2022hyg,Hu:2023mgc}. For the general denominator
it is pointed out in \cite{Guan:2023avw} that  masses could play the role of fugacity variable in generating functions.

Finding the proper organization to define the generating function is only the first step. What is more crucial is to establish its differential equations and then solve them analytically.  In \cite{Feng:2022hyg}, differential equations have been established using the PV-reduction method. When using the same idea to treat two-loop integrals,  we could not get enough
differential equations to determine the generating function completely. It is not so surprising, since compared to other methods the PV-reduction method utilizes the least information about loop integrals, i.e., only the tensor structure. Thus when encountering irreducible scalar products in higher loop integrals, the PV-reduction method faces a serious challenge as demonstrated in \cite{Feng:2022iuc}, where the analytic reduction results of sub-one-loop integrals must be
used to solve the two-loop integral reduction. In fact, the formulation of the differential equation for generating function in \cite{Hu:2023mgc}
is grounded on the observation of such sub-one-loop reduction. This result was initially conjectured in \cite{Feng:2022iuc} and subsequently proved in \cite{Chen:2022jux} and \cite{Li:2022cbx} employing either module intersection method or the projective space language associated with their Feynman parametrization.

In contrast to the PV-reduction method, the IBP method
is considered to be a complete method, i.e., it can reduce any integral to basis regardless of the number of loops or the complexity of integrands. Therefore we believe that by using the IBP method we can get sufficient differential equations for the generating function for any loop level.  To verify this belief,  we will restudy generating functions of one-loop integrals in this paper. We will demonstrate that
using the IBP method, we can derive a much simpler differential equation for the generating function of tensor reduction compared to the results in \cite{Feng:2022hyg}.
Furthermore, the differential equation for the generating function of general denominators could also be easily written down and solved. These results will provide valuable guidance when we explore generating functions for higher loop integrals.

The structure of the paper is as follows. In section two, we set up the general framework for generating functions. In section three,
we establish corresponding differential equations using the IBP method. In section four, we resolve the generating function
for tensor reduction, using either the form of series expansion or hypergeometric function.  In section five, we solve the matrix ${\cal H}$, which
represents the reduction of scalar integrals with only a single propagator with power two.
In section six, we address the generating function for an arbitrary denominator. Finally, a brief conclusion and discussion are presented in
section seven. For self-sufficiency, some technical details are presented in the Appendix.


\section{The setup}\label{sec2}

In this paper, we will discuss the generating function for the general one-loop integrals, i.e., with arbitrary numerator and
denominator. In this section, we will set up the foundation of our discussion.

We start from the generating function of tensor reduction of $(n+1)$-gon, which is defined as\footnote{ In this paper, we consider the generating function in the momentum space. In \cite{Bern:1992em}, by using the Feynman parameter form of Feynman integrals,  tensor integrals become integrals with 
	Feynman parameters inserted at the numerator, which can be obtained by appropriately differentiating a scalar integral as shown in Equation (30) in \cite{Bern:1992em}. Thus the scalar integral given in Equation (29) plays the role of generating function. 
}
\be
I_{n+1}^{R}\equiv \int {d^D\ell\over i\pi^{D/2}}{e^{2\ell\cdot R}\over \prod_{i=0}^n ((\ell-K_i)^2-M_i^2)}=\bm{J}_{n+1}\cdot \vec{\a}^{gen}(R)~,~~\label{gen-1-1}
\ee
where the $\bm{J}_{n+1}$ is the row vector of the scalar basis and the $\vec{\a}^{gen}$ is the column vector of generating functions of tensor reduction. In \eref{gen-1-1}, the power of propagator is one, i.e., we consider only the tensor reduction at this moment. 
To discuss the reduction with higher power of propagators, we will use, for example, \eref{gen-a-1}. 

Before going on, let us discuss the scalar basis $\bm{J}_{n+1}$ presented in \eref{gen-1-1}.
In this paper, we will use the IBP identities to establish differential equations for generating functions. A default assumption 
when applying the IBP method is that there are no kinematic relations between external momenta $K_i$
for general momentum configurations. This is true when and only when the spacetime dimension $d\geq n$ (by dimensional regularization,
the $D$ in \eref{gen-1-1} is given by $D=d-2\epsilon$). When this condition is not satisfied, extra labor must be taken because
some integrals are not basis anymore. For example, with $D=4-2\epsilon$, the scalar hexagon integral is not a basis and it 
can be decomposed as the linear combinations of six pentagons. This
problem has been carefully addressed in \cite{sca-red} when the decomposition of any scalar $n$-gon integrals has been worked out
(an intuitive example is also presented in Appendix \ref{sca-red} for reader convenience). This is the limitation\footnote{In \cite{Bern:1992em},
	using the Feynman parametrization form, there is a way to deal with the arbitrary number of propagators. 
} of results in this
paper when applying to phenomenologically interesting integrals with $D=4-2\epsilon$. For $n>4$ we need to use results 
in \cite{sca-red} to write them as the sum of integrals with $n\leq 4$.

There is another issue about the expression \eref{gen-1-1} where we have assigned each propagator with a different mass. It looks to
make a massive overhead since in the real world it is rare to have more than one nonzero mass. However, we want to emphasize that 
although we start with a general mass configuration, it is easy to take the limit to the specific mass configuration. For example,
for a massless propagator, the corresponding basis, i.e., the tadpole, is zero, so we can just neglect the computation of corresponding 
reduction coefficients. When two propagators have the same mass, their corresponding tadpole basis is, in fact, the same. Then we can 
take just one as the independent basis, but the corresponding reduction coefficient will be the sum of these two reduction coefficients.
Another example is that one-mass triangle can be written as a bubble with some coefficient. Then we can remove the triangle from 
the basis $\bm{J}_{n+1}$ and add its reduction coefficients with proper factor to the bubble. To summarize, the above discussions
tell us when applying generating functions obtained in this paper, one can simply set the mass to zero or a particular nonzero number
from the beginning.

With the above clarification, we define the scalar basis as 
\be J_{n+1;\WH {S}}=\int {d^D\ell\over i\pi^{D/2}}{1\over \prod_{i=0, i\not\in S}^n ((\ell-K_i)^2-M_i^2)},~~~~~~\label{gen-scalar-J}\ee
where $\WH {S}$ is the list of removed propagators, then the row vector $\bm{J}_{n+1}$ will be ordered as
\be {\bm J}=\{ J_{n+1}; J_{n+1;\WH 0},J_{n+1;\WH 1},...,J_{n+1;\WH n} ; J_{n+1; \WH {0  1}},J_{n+1; \WH {02}},...,
J_{n+1; \WH{(n-1)n} };...; J_{n+1; \WH{01...(n-1)}},..\}~.~~~\label{gen-T-7-1}\ee
Similarly, we define the generating functions
\be
I_{n+1;\WH{S}}^{R}\equiv \int {d^D\ell\over i\pi^{D/2}}{e^{2\ell\cdot R}\over \prod_{i=0, i\not\in S}^n ((\ell-K_i)^2-M_i^2)}=\bm{J}_{n+1}\cdot \vec{\a}^{gen}_{n+1;\WH{S}}~.~~\label{gen-1-1-1}
\ee
There are two remarks regarding  \eref{gen-1-1-1}. First, according to our iterative strategy, since $I_{n+1;\WH{S}}^{R}$ has fewer number of propagators, when solving
$\vec{\a}^{gen}$ defined in \eref{gen-1-1}, we  treat $\vec{\a}^{gen}_{n+1;\WH{S}}$ as known functions. Secondly, the right hand side
of \eref{gen-1-1-1} has been written in  the expansion of larger basis  $\bm{J}_{n+1}$
instead of $\bm{J}_{n+1;\WH S}$. Thus the column vector $\vec{\a}^{gen}_{n+1;\WH{S}}$
should be extended also, i.e., the component of vector $\vec{\a}^{gen}_{n+1;\WH{S}}$ should be written as  $(\vec{\a}^{gen}_{n+1;\WH{S}})_{\WH{S}_1}$ by our convention given in \eref{gen-T-7-1}. One obvious result from the extension is that  $(\vec{\a}^{gen}_{n+1;\WH{S}})_{\WH{S}_1}=0$ as long as $S\not\subseteq S_1$.

Next, we consider a little bit more general generating functions defined by
\be
I_{n+1;\{\bm{a}\}}^{R}\equiv \int {d^D\ell\over i\pi^{D/2}}{e^{2\ell\cdot R}\over \prod_{i=0}^n ((\ell-K_i)^2-M_i^2)^{a_i}}=\bm{J}_{n+1}\cdot \vec{\a}^{gen}_{\bm{a}}~~~\label{gen-1-2}
\ee
where $\bm{a}$ is the vector of powers of propagators with $a_i\geq 1,\forall i$. One interesting application of such general integrals
is the construction of a uniform transcendental basis. There are two different approaches  to the problem \eref{gen-1-2}.
The first approach is based on  the observation
\be I^{R}_{n+1;\{\bm{a}\}}= \left(\prod_{j=0}^n {1\over (a_j-1)!}{\d^{a_j-1}\over \d (M_j^2)^{a_j-1}}\right)I^{R}_{n+1}~.~~\label{gen-1-3}\ee
Using the result
\be {\d\over \d (M_j^2)}I^{R}_{n+1} =  \left({\d\over \d (M_j^2)}\bm{J}_{n+1}\right)\cdot \vec{\a}^{gen}+\bm{J}_{n+1}\cdot \left({\d\over \d (M_j^2)}\vec{\a}^{gen}\right)
\equiv  \bm{J}_{n+1}\cdot {\cal D}_{n+1;j}\vec{\a}^{gen}~,~~\label{gen-1-4}\ee
where we have rewritten  the differential action over basis as
\be {\d\over \d (M_j^2)}\bm{J}_{n+1}\equiv \bm{J}_{n+1}{\cal H}_{n+1;j}~~~\label{gen-1-5}\ee
 and  defined the
"covariant derivative" as
\be {\cal D}_{n+1;j}\equiv {\d\over \d (M_j^2)}+ {\cal H}_{n+1;j}~,~~\label{gen-1-6}\ee
we can obtain the reduction of general integrals in \eref{gen-1-2} as
\be I^{R}_{n+1;\{\bm{a}\}}= \bm{J}_{n+1}\cdot \left\{ \left(\prod_{j=0}^n {1\over (a_j-1)!}{\cal D}^{a_j-1}_{n+1;j}\right) \vec{\a}^{gen}\right\}~.~~\label{gen-1-8b}\ee
Thus to completely solve the problem by the first approach, we just need to find the matrix ${\cal H}_{n+1;j}$. Before presenting the
systematical algorithm to write it down in a later section,  we give some general discussions:
\begin{itemize}
\item (a) Since the differential operators ${\d\over \d (M_j^2)}$ are commutative with  each other on the left-hand side of \eref{gen-1-4}, we have following
"integrability conditions" as
\be [{\cal D}_{n+1;i}, {\cal D}_{n+1;j}]=0,~~~\forall i,j~,~~\label{gen-1-8-1a}\ee
which is equivalent to
\be \left({\d\over \d (M_i^2)}  {\cal H}_{n+1;j}\right)+ {\cal H}_{n+1;i} {\cal H}_{n+1;j}-
\left({\d\over \d (M_j^2)}  {\cal H}^{n+1}_i\right)    -{\cal H}_{n+1;j} {\cal H}_{n+1;i}=0~.~~\label{gen-1-8-1b}\ee
Relations \eref{gen-1-8-1b} will give a strong consistent check for matrices ${\cal H}_{n+1;j}$.

\item (b) Elements of the matrix can be represented as $({\cal H}_{n+1;j})_{\WH{S}_1,\WH{S}_2 }$ where the list $S_i$ indicates the removal of propagators from the $(n+1)$-gon (notably, for the first
column $S_2=\emptyset$). Thus according to the definition \eref{gen-1-5}, The $\WH S_2$-th column of the matrix  $({\cal H}_{n+1;j})$ represents  the reduction of scalar integral obtained from
$J_{n+1; \WH {S}_2}$
by increasing the power of
$j$-th propagator from one to two. With this picture, we  immediately obtain the following results:
  \begin{itemize}
  	\item When  $j\in S_2$, $({\cal H}_{n+1;j})_{\WH{S}_1,\WH{S}_2 }=0, \forall S_1$,  since ${\d\over \d M_j^2}J_{n+1; \WH{S}_2}=0$.
  	
  	\item When $S_2\not\subseteq S_1$ ($\emptyset \subseteq S$ for any list $S$), $({\cal H}_{n+1;j})_{\WH{S}_1,\WH{S}_2 }=0$
  	 since the reduction of ${\d\over \d M_j^2}J_{n+1; \WH{S}_2}$ can not reach the scalar basis $J_{n+1; \WH{S}_1}$.

  \end{itemize}

\item (c)   By our iterative construction, only the first column of the matrix ${\cal H}_{n+1;j}$ is unknown, since all
other columns are the reduction of subsectors (lower gons), after naturally embedded to the larger basis $\bm{J}_{n+1}$ by the exactly
same remark after the equation \eref{gen-1-1-1}.

\end{itemize}

The second approach to solve \eref{gen-1-2} is to  sum over all $\bm{a}$ as
\be \sum_{a_0,\cdots,a_n=1}^\infty t_0^{a_0-1}\cdot t_1^{a_1-1}\cdot t_2^{a_2-1}\cdot \cdots \cdot t_n^{a_n-1}I_{n+1;\{\bm{a}\}}^{R}\ee
to reach
\be
I_{n+1}^{R}(\bm{t})\equiv \int {d^D\ell\over i\pi^{D/2}}{e^{2\ell\cdot R}\over \prod_{i=0}^n ((\ell-K_i)^2-M_i^2-t_i)}~.~~\label{gen-a-1}
\ee
In fact,  $I_{n+1}^{R}(\bm{t})$ is the {\it most  general generating function} for one loop reduction we are looking for\footnote{We should remember the limitation caused by the spacetime dimension by our IBP method. }. It
gives the reduction coefficients of one loop integrals with arbitrary numerator and denominator
after the expansion of $R, \bm{t}$.

The expression \eref{gen-a-1} is very similar to  \eref{gen-1-1} except for the mass shifting  $M_i^2\to M_i^2+t_i$. Thus using  \eref{gen-1-1} we have immediately
\bea I_{n+1}^{R}(\bm{t})=\bm{J}_{n+1}(\bm{t})\cdot \vec{\a}^{gen}(R,\bm{t})~~~\label{gen-a-2}\eea
where the scalar basis $\bm{J}_{n+1}(\bm{t})$ is the one in \eref{gen-scalar-J} with the mass shifting $M_i^2\to M_i^2+t_i$.
To complete the reduction \eref{gen-a-2}, we need to find the reduction
\be \bm{J}_{n+1}(\bm{t})=\bm{J}_{n+1}~\cdot {\cal G}(\bm{t}) ~~~\label{gen-a-2-1}\ee
where the matrix ${\cal G}$ depends on $\bm{t}$ only, but not $R$. In other words, to compute ${\cal G}(\bm{t})$, we can start from  \eref{gen-a-1} by setting $R=0$. Assembling all together we finally have
\be I_{n+1}^{R}(\bm{t})=\bm{J}_{n+1}\cdot \vec{\gamma}^{gg}(R,\bm{t}),~~~~~~~~~ \vec{\gamma}^{gg}(R,\bm{t})\equiv {\cal G}(\bm{t})   \vec{\a}^{gen}(R,\bm{t})~.~~\label{gen-a-3}\ee
Although $\vec{\gamma}^{gg}(R,\bm{t})$ is the most general generating function for the reduction,
it is more complicated comparing to $\vec{\a}^{gen}$ in \eref{gen-1-1} with the appearance of multiple variables $t_i$. The result
\eref{gen-a-3} shows that its computation can be reduced to two simpler computations of generating functions ${\cal G}(\bm{t})$ and   
$\vec{\a}^{gen}(R,\bm{t})$. In fact, by comparing \eref{gen-1-5} and \eref{gen-a-2-1} one can see that 
\bea {\cal H}_{n+1;j}=\left\{{\partial \over \partial t_j}{\cal G}(\bm{t})\right\}_{\bm{t}\to 0}~~\label{gen-a-4}\eea

The above two approaches are, in fact, related to each other as follows. Using \eref{gen-1-8b} one finds
\bea & & \sum_{a_0,\cdots,a_n=1}^\infty {t_0^{a_0-1}\cdot t_1^{a_1-1}\cdot \cdots \cdot t_n^{a_n-1}}I^{R}_{n+1;\{\bm{a}\}}=  \prod_{j=0}^n \left(\sum_{a_j=1}^\infty {t_j^{a_j-1}\over (a_j-1)!}\left({\d\over \d M_j^2}\right)^{a_j-1}\right)I^{R}_{n+1}\nn
& = & \bm{J}_{n+1}\left(  \prod_{j=0}^n \left(\sum_{a_j=1}^\infty {t_j^{a_j-1}\over (a_j-1)!}\left({\cal D}_{n+1;j}\right)^{a_j-1}\right)\vec{\a}^{gen}\right)=\bm{J}_{n+1}\cdot \left(e^{\sum_{j=0}^n t_j{\cal D}_{n+1;j}}\vec{\a}^{gen}\right)~~~\label{gen-12-1}\eea
thus one gets
\be \vec{\gamma}^{gg}(R,\bm{t})=e^{\sum_{j=0}^n t_j{\cal D}_{n+1;j}}\vec{\a}^{gen}(R)~~~\label{gen-12-2}\ee
Although knowing the matrices ${\cal H}_{n+1;j}$ is sufficient to compute $\vec{\gamma}^{gg}(R,\bm{t})$ via equation  \eref{gen-12-2},
the evaluation of differential action $e^{\sum_{j=0}^n t_j{\cal D}_{n+1;j}}$ is still a non-trivial task. For the completion of the paper, we will systematically and  analytically solve $\vec{\a}^{gen}(R)$, ${\cal H}_{n+1;j}$ and ${\cal G}(\bm{t})$ one by one in subsequent  sections. In fact, the computation of ${\cal G}(\bm{t})$ gives another simple way to computate the matrix ${\cal H}_{n+1;j}$
as given by \eref{gen-a-4}.

Before ending this section, it is important to emphasize that explicit expressions of generating functions in \eref{gen-1-1}, \eref{gen-1-2} and \eref{gen-a-1} rely on the choice of loop momentum $\ell$. For the convenience of subsequent discussions, unless explicitly stated, we adopt the convention that $K_0=0$.
\section{The IBP relations}

In this section, we will derive differential equations for generating function \eref{gen-1-1} coming from  IBP relations. The results presented in this section form the foundation for all discussions in later sections.

\subsection{Fundamental IBP relations}

For fundamental IBP relations, the first one is
\bea 0 & = & \int {d^D\ell\over i\pi^{D/2}}{d\over d \ell^\mu}\left\{ \ell^\mu {e^{2\ell\cdot R}\over (\ell^2-M_0^2)\prod_{i=1}^n ((\ell-K_i)^2-M_i^2)}\right\}\nn
& = & \int {d^D\ell\over i\pi^{D/2}} {D e^{2\ell\cdot R}\over (\ell^2-M_0^2)\prod_{i=1}^n ((\ell-K_i)^2-M_i^2)}
+\int {d^D\ell\over i\pi^{D/2}} {2\ell\cdot R e^{2\ell\cdot R}\over (\ell^2-M_0^2)\prod_{i=1}^n ((\ell-K_i)^2-M_i^2)}\nn
& & + \int {d^D\ell\over i\pi^{D/2}} {-2\ell^2 e^{2\ell\cdot R}\over (\ell^2-M_0^2)^2\prod_{i=1}^n ((\ell-K_i)^2-M_i^2)}\nn
& & +\sum_{j=1}^n \int {d^D\ell\over i\pi^{D/2}} {-2\ell\cdot (\ell-K_j)  e^{2\ell\cdot R}\over ((\ell-K_j)^2-M_j^2)^2(\ell^2-M_0^2)\prod_{i=1,i\neq j}^n ((\ell-K_i)^2-M_i^2)}~.~~\label{gen-IBP-1-1}
\eea
Among these four terms at the right-hand side of \eref{gen-IBP-1-1}, the first term is just $D \bm{J}_{n+1}\cdot \vec{\a}^{gen}$. The second term is $\left(R\cdot {\d\over \d R}\right)I_{n+1}^{R}= \bm{J}_{n+1}\cdot \left(\left(R\cdot {\d\over \d R}\right)\vec{\a}^{gen}\right)$.
For the third term, after writing $2\ell^2= 2((\ell^2-M_0^2)+M_0^2)$ we get
	\be \int {d^D\ell\over i\pi^{D/2}} {-2M_0^2 e^{2\ell\cdot R}\over (\ell^2-M_0^2)^2\prod_{i=1}^n ((\ell-K_i)^2-M_i^2)} +\int {d^D\ell\over i\pi^{D/2}} {-2 e^{2\ell\cdot R}\over (\ell^2-M_0^2)\prod_{i=1}^n ((\ell-K_i)^2-M_i^2)} ~.~~\label{gen-IBP-1-2}\ee
By discussions from \eref{gen-1-2} to \eref{gen-1-8b}, the first term
is just $\bm{J}_{n+1}\cdot (-2M_0^2{\cal D}_{n+1;0}\vec{\a}^{gen})$. To simplify the notation, we will define
	\be {\cal D}_{n+1;i}\vec{\a}^{gen}\equiv \vec{X}_i ~~~\label{gen-IBP-1-3}\ee
and the third term of \eref{gen-IBP-1-1} can be written as $\bm{J}_{n+1}\cdot \left(-2M_0^2 \vec{X}_0-2  \vec{\a}^{gen}\right)$.
By similar manipulations, the fourth term of \eref{gen-IBP-1-1} is
	\be \bm{J}_{n+1}\cdot \left(\sum_{j=1}^n f_j \vec{X}_j- \vec{\a}^{gen}-{\cal D}_{n+1;j} \vec{\a}_{n+1;\WH 0}^{gen}\right)~~~\label{gen-1-10}\ee
	where $\vec{\a}_{n+1;\WH 0}^{gen}$ is known as defined in \eref{gen-1-1-1} and
$ f_j\equiv K_j^2-M_j^2-M_0^2$.
	It is worth to emphasize that although naively ${\cal D}_{n+1;j}$ contains the unknown matrix ${\cal H}_{n+1;j}$, the
	entire combination ${\cal D}_{n+1;j} \vec{\a}_{n+1;\WH 0}^{gen}$ is, in fact, known due to the following observation
	\be {\d\over \d (M_j^2)}I^{R}_{n+1;\WH 0}=\bm{J}_{n+1}\cdot \left( {\cal D}_{n+1;j} \vec{\a}_{n+1;\WH 0}^{gen}\right)~~~\label{gen-1-9}\ee
	with our iterative assumption.
	Putting all together, the IBP relation \eref{gen-IBP-1-1} becomes
\be 0 = \bm{J}_{n+1}\cdot \left\{\left(D-2-n+R\cdot {d\over d R}\right) \vec{\a}^{gen}-2M_0^2 \vec{X}_0 +\sum_{j=1}^n \left( f_j \vec{X}_j-{\cal D}_{n+1;j} \vec{\a}_{n+1;\WH 0}^{gen}\right)\right\}~~~~~~\label{gen-1-11}\ee
From it we read out the differential equation for the generating function
\be 0 =  \left\{\left(D-2-n+R\cdot {d\over d R}\right) \vec{\a}^{gen}-2M_0^2 \vec{X}_0 +\sum_{j=1}^n \left( f_j \vec{X}_j-{\cal D}_{n+1;j} \vec{\a}_{n+1;\WH 0}^{gen}\right)\right\}~~~~~~\label{gen-1-11-1}\ee

Having worked through the IBP relation \eref{gen-IBP-1-1}, we merely write down other fundamental IBP relations and
corresponding differential equations without details. From IBP relations\footnote{In fact, instead of using ${d\over d \ell^\mu} K_r^\mu$, it is better to
	use ${d\over d \ell^\mu}(\ell- K_r)^\mu$ as shown in \cite{Duplancic:2003tv} (see also \cite{Guan:2023avw}), which is the origin of the simplification used in \eref{gen-T-2-4}.}
\be 0  =  \int {d^D\ell\over i\pi^{D/2}}{d\over d \ell^\mu}\left\{ K_r^\mu {e^{2\ell\cdot R}\over (\ell^2-M_0^2)\prod_{i=1}^n ((\ell-K_i)^2-M_i^2)}\right\}~~~\label{gen-IBP-2-1}
\ee
with $r=1,...,n$, we have
\bea 0 & = & \left\{(2K_r\cdot R) \vec{\a}^{gen}+ \left( {\cal D}_{n+1;0}\vec{\a}_{n+1;\WH r}^{gen}\right)
-g_r \vec{X}_0- \left( {\cal D}_{n+1;r}\vec{\a}_{n+1;\WH 0}^{gen}\right)+(-g_r+2K^2_r)\vec{X}_r\right.\nn
& & \left. +\sum_{j=1,j\neq r}^n\left( {\cal D}_{n+1;j}\vec{\a}_{n+1;\WH r}^{gen}\right)-\left( {\cal D}_{n+1;j}\vec{\a}_{n+1;\WH 0}^{gen}\right)+(-g_r+2K_r\cdot K_j)\vec{X}_j\right\}  ~~~\label{gen-IBP-2-10}\eea
with $ g_r=K_r^2-M_r^2+M_0^2$.
From IBP relation
\be 0  =  \int {d^D\ell\over i\pi^{D/2}}{d\over d \ell^\mu}\left\{ R^\mu {e^{2\ell\cdot R}\over (\ell^2-M_0^2)\prod_{i=1}^n ((\ell-K_i)^2-M_i^2)}\right\}~~~\label{gen-IBP-3-1}
\ee
we have
\be 0
 =  2R^2 \vec{\a}^{gen}+\sum_{j=0}^n \left(\left(2R\cdot K_j-R\cdot {\d\over \d R} \right) \vec{X}_j\right)~~~\label{gen-IBP-3-5} \ee
where  the $K_0=0$ convention in \eref{gen-IBP-3-5} should be remembered. Compared to the familiar IBP relations
\eref{gen-IBP-1-1} and \eref{gen-IBP-2-1},
the \eref{gen-IBP-3-1} is a new one by the introduction of the auxiliary vector $R$. It will play a crucial role
when solving the generating function  $\vec{\a}^{gen}$.

\subsection{Simplifying differential equations}

Now we try to simplify the differential equations \eref{gen-1-11-1}, \eref{gen-IBP-2-10} and \eref{gen-IBP-3-5}.  With definitions
\bea \vec{Y}_0 &\equiv & \left(D-2-n+ R\cdot {\d\over \d R}\right) \vec{\a}^{gen} -\sum_{j=1}^n {\cal D}_{n+1;j} \vec{\a}_{n+1;\WH 0}^{gen}~~~~~~\label{gen-1-12}\\
\vec{Y}_r & \equiv & (2K_r\cdot R) \vec{\a}^{gen}+ \left( {\cal D}_{n+1;0}\vec{\a}_{n+1;\WH r}^{gen}\right)
- \left( {\cal D}_{n+1;r}\vec{\a}_{n+1;\WH 0}^{gen}\right)\nn & & +\sum_{j=1,j\neq r}^n\left( {\cal D}_{n+1;j}\vec{\a}_{n+1;\WH r}^{gen}\right)-\left( {\cal D}_{n+1;j}\vec{\a}_{n+1;\WH 0}^{gen}\right) ~~~\label{gen-IBP-2-12}\eea
differential equations \eref{gen-1-11-1} and \eref{gen-IBP-2-10} can be cast to
\bea \vec{Y}_0 & = & 2M_0^2 \vec{X}_0 -\sum_{j=1}^n  (K_j^2-M_j^2-M_0^2) \vec{X}_j~~~~~\label{gen-T-1-1}\\
\vec{Y}_r & = & (K_r^2-M_r^2+M_0^2) \vec{X}_0-(-(K_r^2-M_r^2+M_0^2)+2K^2_r)\vec{X}_r\nn & & -\sum_{j=1,j\neq r}^n (-(K_r^2-M_r^2+M_0^2)+2K_r\cdot K_j)\vec{X}_j  ~~~\label{gen-T-1-2}\eea
With the definition of  the {\bf modified Caylay matrix $F$} with  elements
\be f_{ij}=(K_i-K_j)^2-M_i^2-M_j^2,~~~~i,j=0,...,n;~~~K_0=0~~~~~\label{gen-T-2-2}\ee
and
\be \vec{\cal Y}_r\equiv \vec{Y}_r-\vec{Y}_0,~~~~\vec{\cal Y}_0\equiv-\vec{Y}_0~,~~~~\label{gen-T-2-4}\ee
 \eref{gen-T-1-1} and \eref{gen-T-1-2} can be written beautifully as\footnote{Above simplification is motivated by Eq(16) of \cite{Guan:2023avw} (see also \cite{Duplancic:2003tv})}
\be \vec{\cal Y}_r =  \sum_{j=0}^n f_{rj} \vec{X}_j~,~~~r=0,1...,n~~~~\label{gen-T-2-3}\ee
One immediately solves\footnote{As we have carefully discussed  before, the existence of $F^{-1}$ puts the condition that $d\geq n$, which is the limit of using only IBP relations. }
\be \boxed{ \vec{X}_i= (F^{-1})_{ij}\vec{\cal Y}_j}~~~~~\label{gen-T-3-3}\ee
where $\vec{\cal Y}_r$, $r=0,1,...,n$ can be simplified as
\be \vec{\cal Y}_r =  -\left(D-2-n-2K_r\cdot R+ R\cdot {\d\over \d R}\right) \vec{\a}^{gen}+\sum_{j=0,j\neq r}^n\left( {\cal D}_{n+1;j}\vec{\a}_{n+1;\WH r}^{gen}\right)~~~~~\label{gen-T-3-2}\ee
by using  \eref{gen-1-12} and \eref{gen-IBP-2-12}.

Now we put \eref{gen-T-3-3} back to \eref{gen-IBP-3-5} to get
\bea \boxed {0  =  2R^2 \vec{\a}^{gen}+\sum_{j=0}^n \left(\left(2R\cdot K_j-R\cdot {\d\over \d R} \right) (F^{-1})_{jr} \vec{\cal Y}_r\right)}~~~\label{gen-IBP-3-5-1} \eea
Equations \eref{gen-IBP-3-5-1} and \eref{gen-T-3-3} are two main results of the section. Using \eref{gen-IBP-3-5-1} we can solve
$\vec{\a}^{gen}$ and  using \eref{gen-T-3-3} we can solve the unknown matrix ${\cal H}_{n+1;j}$.
Before ending this section, let us remark that so far  we have only used  fundamental IBP relations  for differential equations.
In fact, one can consider other IBP relations as done in \cite{Kosower:2018obg, Hu:2023mgc}. We will give more
discussions of the issue in the conclusion.

\section{Solving $\vec{\a}^{gen}$}

In this section, we will use differential equations presented in the previous section to solve the generating function $\vec{\a}^{gen}$. The same generating function has been studied in \cite{Feng:2022hyg}. However,  differential equations in \cite{Feng:2022hyg} are established using the PV-reduction method,
thus they are $(n+1)$ partial differential equations. On the contrary, as we will show in this section, using the IBP method a single second-order ordinary differential equation for generating function
can be established, thus solving it is much simpler.

\subsection{Solving by series expansion}

First, we solve it using series expansion.
To simplify notations, let us define
\be B_r=\sum_{j=0,j\neq r}^n\left( {\cal D}_{n+1;j}\vec{\a}_{n+1;\WH r}^{gen}\right)~~~~\label{gen-T-7-5}\ee
which is known by our iterative construction (see the explanation \eref{gen-1-9}). In fact, using the definition \eref{gen-IBP-1-3}
each term at the right hand side of \eref{gen-T-7-5} can be solved  by \eref{gen-T-3-3} with reduced modified Caylay matrix
and the generating function of  $n$-gon.
Then Eq \eref{gen-IBP-3-5-1} can be rearranged to
\bea & & \left\{ 2R^2 -\sum_{j=0}^n \left(2R\cdot K_j-R\cdot {\d\over \d R} \right) (F^{-1})_{jr} \left(D-2-n-2K_r\cdot R+ R\cdot {\d\over \d R}\right) \right\}\vec{\a}^{gen}\nn
& = & - \sum_{j=0}^n \left(2R\cdot K_j-R\cdot {\d\over \d R} \right) (F^{-1})_{jr} B_r~.~~\label{gen-IBP-3-5-1-2}\eea
Now we expand
\bea \vec{\a}^{gen} & = & \sum_{i_0,...,i_n=0}^\infty\vec{\a}_{i_0 i_1...i_n} (R^2)^{i_0} (2K_1\cdot R)^{i_1}...(2K_n\cdot R)^{i_n}\nn
B_r & = & \sum_{i_0,...,i_n=0}^\infty\vec{b}_{r;i_0 i_1...i_n} (R^2)^{i_0} (2K_1\cdot R)^{i_1}...(2K_n\cdot R)^{i_n}~~~\label{gen-IBP-3-5-2} \eea
Putting \eref{gen-IBP-3-5-2} to \eref{gen-IBP-3-5-1-2} and matching coefficients of independent tensor structures at both
sides, we can solve
\bea & & \vec{\a}_{i_0 i_1...i_n}(2i_0+\sum_{t=1}^n i_t)(D-2-n+(2i_0+\sum_{t=1}^n i_t) )\sum_{j,r=0}^n (F^{-1})_{jr} \nn
& = & -\left[ \sum_{j=1}^n\sum_{r=0}^n (F^{-1})_{jr} \vec{b}_{r;i_0 i_1...(i_j-1)...i_n}- (2i_0+\sum_{t=1}^n i_t)\sum_{j,r=0}^n (F^{-1})_{jr}\vec{b}_{r;i_0 i_1...i_n} \right]\nn
& & +\sum_{j=0}^n\sum_{r=1}^n(F^{-1})_{jr} \vec{\a}_{i_0 i_1...(i_r-1)...i_n}(2i_0+\sum_{t=1}^n i_t)\nn
& &  +\sum_{j=1}^n\sum_{r=0}^n(F^{-1})_{jr}\vec{\a}_{i_0 i_1...(i_j-1)...i_n}(D-2-n+2i_0+\sum_{t=1}^n i_t-1 )\nn
& & - \sum_{j,r=1}^n (F^{-1})_{jr}\vec{\a}_{i_0 i_1...(i_j-1)...(i_r-1)...i_n}-2\vec{\a}_{(i_0-1) i_1...i_n}~~~\label{gen-IBP-3-5-10}\eea
If we consider the symmetry of matrix $F$, it can be simplified as
\bea & & \vec{\a}_{i_0 i_1...i_n}(2i_0+\sum_{t=1}^n i_t)(D-2-n+(2i_0+\sum_{t=1}^n i_t) )\sum_{j,r=0}^n (F^{-1})_{jr} \nn
& = & -\left[ \sum_{j=1}^n\sum_{r=0}^n (F^{-1})_{jr} \vec{b}_{r;i_0 i_1...(i_j-1)...i_n}- (2i_0+\sum_{t=1}^n i_t)\sum_{j,r=0}^n (F^{-1})_{jr}\vec{b}_{r;i_0 i_1...i_n} \right]\nn
& &  +\sum_{j=1}^n\sum_{r=0}^n(F^{-1})_{jr}\vec{\a}_{i_0 i_1...(i_j-1)...i_n}(D-3-n+4i_0+2\sum_{t=1}^n i_t)\nn
& & - \sum_{j,r=1}^n (F^{-1})_{jr}\vec{\a}_{i_0 i_1...(i_j-1)...(i_r-1)...i_n}-2\vec{\a}_{(i_0-1) i_1...i_n}~~~\label{gen-IBP-3-5-10b}\eea
There are a few remarks for the results \eref{gen-IBP-3-5-10} and \eref{gen-IBP-3-5-10b}. First,  we make the convention that
coefficients $(\a/b_r)_{i_0...i_n}=0$ if any subscript $i_k<0$.
 Secondly, for the first term in the last line of \eref{gen-IBP-3-5-10} and \eref{gen-IBP-3-5-10b}, if $j=r$, we should interpret
	$\vec{\a}_{i_0 i_1...(i_j-1)...(i_r-1)...i_n}$ as $ \vec{\a}_{i_0 i_1...(i_j-2)...i_n}$.
Thirdly since for $\vec{\a}$, any subscript at the right-hand side of \eref{gen-IBP-3-5-10} and \eref{gen-IBP-3-5-10b} is always
	equal or at least one less than  those at the left-hand side, we can solve $\vec{\a}_{i_0 i_1...i_n}$
	recursively using \eref{gen-IBP-3-5-10b} with the obviously initial condition
	$\vec{\a}_{00...0}=\{1,0,0,...,0\}$.
Fourthly  for $n\geq 1$, the summation is\footnote{see Eq(B.6) in \cite{Abreu:2017ptx}}
	\be \sum_{i,j=0}^n  (F^{-1})_{i,j}= (-)^n 2^n {{\rm Gram}\over {\rm det}(F)},~~~\label{gen-T-2-2-1a}\ee
	where the ${\rm Gram}$ is defined as
	\be {\rm Gram}={\rm det}((K_a-K_*)\cdot (K_b-K_*)),~~~ a,b\in {\cal K}\setminus \{K_*\}~~~\label{gen-T-2-2-1b}\ee
	with $K_*$ any fixed moment in the set ${\cal K}=\{K_0,K_1,...,K_n\}$. For our convention, $K_0=0$ so it
	is convenient to take $K_*=K_0$. Using \eref{gen-T-2-2-1a} one can simplify the calculation of \eref{gen-IBP-3-5-10} and \eref{gen-IBP-3-5-10b}.
Fifthly when we consider the first component of the vector $\vec{\a}$, i.e., $(\vec{\a})_{\emptyset}$ , since all contributions from $\vec{b}$ is zero,
	\eref{gen-IBP-3-5-10b} can be further simplified to
	\bea & & (\vec{\a}_{i_0 i_1...i_n})_{\emptyset}(2i_0+\sum_{t=1}^n i_t)(D-2-n+(2i_0+\sum_{t=1}^n i_t) )\sum_{j,r=0}^n (F^{-1})_{jr} \nn
	& = & \sum_{j=1}^n\sum_{r=0}^n(F^{-1})_{jr}(\vec{\a}_{i_0 i_1...(i_j-1)...i_n})_{\emptyset}(D-3-n+4i_0+2\sum_{t=1}^n i_t)\nn
	& & - \sum_{j,r=1}^n (F^{-1})_{jr}(\vec{\a}_{i_0 i_1...(i_j-1)...(i_r-1)...i_n})_{\emptyset}-2(\vec{\a}_{(i_0-1) i_1...i_n})_{\emptyset}~.~~\label{gen-IBP-3-5-10-First}\eea
 It gives the reduction coefficients from
	$(n+1)$-gon to $(n+1)$-gon.

Result \eref{gen-IBP-3-5-10b} is one of the main results in the paper. It is easy to implement in Mathematica and gives 
analytic expressions for reduction coefficients.  Now we use several examples to demonstrate the use of \eref{gen-IBP-3-5-10b}.

\subsubsection{The tadpole}
For this case, since all $B_r=0$, \eref{gen-IBP-3-5-10b} is simplified to
\be  {\a}_{i_0 }(2i_0)(D-2+2i_0 ) F_{00}^{-1}=-2{\a}_{(i_0-1) }~~~\label{Ser-tad-1}\ee
with
\be F_{00}= -2M_0^2~.\ee
Thus it is easy to solve
\bea {\a}_{i_0 }={M_0^2\over i_0 ({D\over 2}+i_0-1)}{\a}_{i_0-1 },~~~{\a}_{0}=1\eea
which reproduces the result given in \cite{Feng:2022hyg}, i.e.,
\be \a^{gen}(M_0,R)=\sum_{n=0}^\infty { ( M_0^2 R^2 )^n\over n!  \left({D\over 2} \right)_n}= ~_0 F_1(\emptyset; {D\over 2}; M_0^2 R^2)~~~~\label{Gen-tad-sol-1-9}\ee
%
where the  {\bf Pochhammer symbol} is defined by\footnote{From the definition one can see that $(x)_{n=0}=1,\forall x$. }
\be (x)_n= {\Gamma(x+n)\over \Gamma(x)}=\prod_{i=1}^n (x+(i-1))~.~~~\label{Poch-1}\ee

Before ending this part, let us remark that  using the momentum shifting, it is easy to see that
\be \int {d^D\ell\over i\pi^{D/2}} {e^{2\ell\cdot R}\over ((\ell-K_1)^2-M_1^2) }= e^{2K_1\cdot R} \a^{gen}(M_1,R)\int {d^D\ell\over i\pi^{D/2}} {1\over ((\ell-K_1)^2-M_1^2) }~~~~\label{Gen-tad-sol-1-10}\ee
which will be useful for later iterative computation.

\subsubsection{The bubble}

Now let us consider the first nontrivial case, i.e., the bubble. According to our convention, the basis is
\be \bm{J}_2=\{ I_2, I_{2;\WH 0}, I_{2;\WH 1}\}~~~~~\label{bub-1-1}\ee
The modified Caylay matrix $F$ is
\bea F= \left( \begin{array}{cc} -2 M_0^2 & K_1^2-M_0^2-M_1^2 \\ K_1^2-M_0^2-M_1^2 & -2 M_1^2 \end{array}\right)~.~~~\label{bub-2-5}\eea
Using \eref{gen-1-9} one can find\footnote{To carry out
	the computation, one need to know the matrix ${\cal H}_{n+1;j}$, which will be discussed in the section \ref{H-matrix}.} for the expansion $B_i
	 =  \sum_{i_0 i_1} \vec{b}_{i;i_0i_1} (R^2)^{i_0}(2K_1\cdot R)^{i_1}$ with
\bea \vec{b}_{0;i_0i_1}= {(D-2+2i_0)\over 2 M_1^2}{1\over i_1!}\left( \begin{array}{c} 0 \\  { ( M_1^2 )^{i_0}\over i_0!  \left({D\over 2} \right)_{i_0}} \\ 0 \end{array}\right)~,~~~\vec{b}_{1;i_0i_1}= {(D-2+2i_0)\over 2 M_0^2}\delta_{i_1,0}\left( \begin{array}{c} 0 \\ 0\\ { ( M_0^2 )^{i_0}\over i_0!  \left({D\over 2} \right)_{i_0}} \end{array}\right)~~~~\label{bub-2-7-1}\eea
Although it is not difficult to write down the analytic expressions using \eref{gen-IBP-3-5-10} or \eref{gen-IBP-3-5-10b}, since
these results have been presented in various works many times (see, for example, \cite{Hu:2021nia} ), here we will just give some numerical checks. 	We choose  numeric values of various kinetic variables as follows
\be M_0^2\to \frac{123}{100},\ M_1^2\to \frac{64}{25},\ K_1^2\to \frac{69}{50},~~
	R^2\to \frac{133}{10},\  K_1\cdot R\to \frac{367}{100},\ D\to \frac{969}{50}.
\ee
Then we have
\bea
{\rm Rank~One}: &~~&  (2R\cdot K_1)\Vec{\alpha}_{0,1}=(0.132971,\ 2.65942,\ -2.65942)^T\nn
{\rm Rank~Two}: &~~& 2!((2R\cdot K_1)^2\Vec{\alpha}_{0,2}+R^2\Vec{\alpha}_{1,0})=(0.964908,\ 20.252,\ -0.346648)^T\nn
{\rm Rank~Three}: &~~&  3!((2R\cdot K_1)^3\Vec{\alpha}_{0,3}+R^2(2R\cdot K_1)\Vec{\alpha}_{1,1})=(0.380212,\ 174.709,\ -13.8036)^T.
\eea
They agree with results by FIRE6 \cite{Smirnov:2019qkx}.

\subsubsection{Triangle}

Again, we give only the numerical check. We choose  numeric values of various kinetic variables as follows:
\begin{align}
	&M_0^2\to \frac{123}{100},\ M_1^2\to \frac{64}{25},\ M_2^2\to \frac{357}{100},\ K_1^2\to \frac{69}{50},\ K_2^2\to \frac{61}{25}\notag\\
	&R^2\to \frac{133}{10},\  K_1\cdot R\to \frac{367}{100},\ K_2\cdot R\to \frac{523}{100},\ K_1\cdot K_2\to \frac{67}{100},\ D\to \frac{969}{50}.
\end{align}
Then we have
\bea
{\rm Rank~One}: &~~&  (2R\cdot K_1)\Vec{\alpha}_{0,1,0}+(2R\cdot K_2)\Vec{\alpha}_{0,0,1}
=(0.256446,\ 3.49834,\ -1.86777,\ -1.63054,\ 0,\ 0,\ 0)^T.\nn
{\rm Rank~Two}: &~~&   2!((2R\cdot K_1)^2\Vec{\alpha}_{0,2,0}+(2R\cdot K_2)^2\Vec{\alpha}_{0,0,2}+(2R\cdot K_1)(2R\cdot K_1)\Vec{\alpha}_{0,1,1}+R^2\Vec{\alpha}_{1,0,0})\nn & &
=(-0.523243,\ 29.5767,\ -0.881583,\ -0.639263,\ 8.33983,\ -6.53695,\ -1.80288)^T.\nn
{\rm Rank~Three}: &~~&  3!((2R\cdot K_1)^3\Vec{\alpha}_{0,3,0}+(2R\cdot K_2)^3\Vec{\alpha}_{0,0,3}+(2R\cdot K_1)^2(2R\cdot K_2)\Vec{\alpha}_{0,2,1}\nn & &
+(2R\cdot K_1)(2R\cdot K_2)^2\Vec{\alpha}_{0,1,2}+R^2(2R\cdot K1)\Vec{\alpha}_{1,1,0}+R^2(2R\cdot K_2)\Vec{\alpha}_{1,0,1})\nn & &
=(-0.43628,\ 260.958,\ 0.723717,\ 0.0778428,\ 3.58409,\ -67.3923,\ 0.605462)^T.
\eea
They agree with results by FIRE6 \cite{Smirnov:2019qkx}.
\subsection{Solving by differential equations}

The equation \eref{gen-IBP-3-5-1-2} is the differential equation for generating function $\vec{\a}^{gen}$ already. However,
it is a partial differential equation, which is harder to solve than the ordinary differential equation in general. Fortunately,
the differential operator on
the left-hand side of \eref{gen-IBP-3-5-1-2} is special and we can transfer it into an ordinary differential equation by
defining a new variable
\be R=t \W R~.~~~\label{gen-IBP-3-diff-1}\ee
With this definition, it is easy to see that for any function $f(R)=\sum f_{i_0...i_n}(R^2)^{i_0} (2K_1\cdot R)^{i_1}...(2K_n\cdot R)^{i_n}\equiv \W f(t,\W R)$
\be  R\cdot {d\over d R}f(R)
=  t{d\over dt}\W f(t,\W R) ~~,~~~  (2K_r \cdot R)f(R)
 =  t  (2K_r \cdot \W R) \W f(t,\W R)~.~~~\label{gen-IBP-3-diff-5}\ee
Using the above results and with some algebraic computations \eref{gen-IBP-3-5-1-2} becomes\footnote{
	A  similar first-order ordinary differential equation has also been given in \cite{Hu:2023mgc}. However, since
	the definition of the generating function is different between \cite{Feng:2022hyg} and \cite{Hu:2023mgc},
	the equation and solution are also different.}
\bea 0 & = & \left[\sum_{j=0}^n \sum_{r=0}^n (F^{-1})_{jr}\right]t {\d^2\over \d t^2}\vec{\W\a}^{gen}(t,\W R)\nn
& & + \left[  (D-2-n+1)\sum_{j=0}^n \sum_{r=0}^n (F^{-1})_{jr}-2\sum_{j=0}^n \sum_{r=1}^n (F^{-1})_{jr}(2K_r\cdot \W R)t\right]{\d\over \d t}\vec{\W\a}^{gen}(t,\W R)\nn
& & + \left[ 2\W R^2 t -(D-2-n+1)\sum_{j=1}^n \sum_{r=0}^n (2\W R\cdot K_j)(F^{-1})_{jr}+\sum_{j=1}^n \sum_{r=1}^n (2\W R\cdot K_j)(F^{-1})_{jr}(2 K_r\cdot \W R)t\right]\vec{\W\a}^{gen}(t,\W R)\nn
& & +\sum_{j=0}^n \left((2\W R\cdot K_j) -{\d\over \d t} \right) (F^{-1})_{jr} \W B_r(t,\W R)~~~~\label{gen-IBP-3-diff-10}\eea
which belongs to the following type of second-order ordinary differential equation
\be At{d^2\over d t^2} W+ (B_0+B_1 t){d\over dt}W+(C_0+C_1 t)W+ {\cal B}(t)=0~.~~~\label{gen-IBP-3-diff-11}\ee
To express coefficients more compactly, let  us define following two row vectors with $(n+1)$ components
\be {\bm I}=(1,1,...,1),~~~{\cal P }=(2K_0\cdot \W R,2K_1\cdot \W R,...,2K_n\cdot \W R )~.~~~\label{gen-IBP-3-diff-11-1}\ee
Using them we have
\bea A &= &  {\bm I}\cdot (F^{-1}) \cdot {\bm I}^T,~~~~~B_0 = (D-(n+1))A\nn
B_1 &= & -2{\bm I}\cdot (F^{-1})\cdot {\cal P}^T= -2{\cal P} \cdot (F^{-1}) \cdot {\bm I}^T,~~~~C_0= {(D-(n+1))\over 2} B_1\nn
C_1 & = & 2\W R^2+{\cal P} \cdot (F^{-1}) \cdot {\cal P}^T~.~~~\label{gen-IBP-3-diff-11-2}\eea
We want to point out that although previous derivation has used the convention $K_0=0$, results \eref{gen-IBP-3-diff-11}, \eref{gen-IBP-3-diff-11-1} and \eref{gen-IBP-3-diff-11-2} can be freely applied to the case $K_0\neq 0$.

The detailed discussion of equation \eref{gen-IBP-3-diff-11}  has been given in the Appendix, where both analytic
expression and series expansion have been given. Using the results \eref{Diff-1-1}, \eref{Diff-2-8} and \eref{Diff-2-9} we write down the solution of \eref{gen-IBP-3-diff-11} as
\bea \vec{\W\a}^{gen}(t,\W R) & = & e^{{-B_1+\sqrt{B_1^2-4 C_1}\over 2}t} (~_1F_1(a_1;b_1;x))\left\{ \vec{c}_1+\int_0^x ds { s^{-b_1} e^s\over (~_1F_1(a_1;b_1;s))^2}\right.\nn & & ~~~~\left.
\left(\vec{c}_2+ \int_0^s d y g(y) y^{b_1-1} ~_1F_1(a_1;b_1;y)e^{-y}\right)\right\}~~~~\label{gen-IBP-3-diff-12}\eea
with
\bea x & = & {-t\sqrt{B_1^2-4C_1 A }\over A},~~~b_1={B_0\over A},~~a_1= {1\over \sqrt{B_1^2-4 C_1 A}}\left( {-B_1+\sqrt{B_1^2-4 C_1 A}\over 2}{B_0\over A}+ {C_0} \right)\nn
\vec{g}(x) & = &  {1\over \sqrt{B_1^2-4A C_1}}e^{{-(B_1-\sqrt{B_1^2-4 AC_1})x\over 2 \sqrt{B_1^2-4 AC_1}}}\vec{\cal B}({-x A\over\sqrt{B_1^2-4C_1 A } })
~~~~\label{gen-IBP-3-diff-13}\eea
Since when $t=0$, $~_1F_1(a_1;b_1;0)=1$, we find $\vec{c}_1= \{1,0,0,...,0\}^T$. The finiteness of ${d \vec{\W\a}^{gen}(t,\W R)\over dt }|_{t=0}$ forces $\vec{c}_2=\vec{0}$.
The result of \eref{gen-IBP-3-diff-12} is given in the vector form. When we consider the first component, i.e.,
the reduction coefficient of $(n+1)$-gon, the generating function is just
\bea (\vec{\W\a}^{gen}(t,\W R))_{\emptyset}=e^{{-B_1+\sqrt{B_1^2-4 C_1 A}\over 2 A} t}~_1F_1(a_1;b_1;x)~~~~\label{gen-IBP-3-diff-14}\eea
Using the analytic result \eref{gen-IBP-3-diff-14} (or the simplified version \eref{gen-IBP-3-diff-14-a2} or \eref{gen-IBP-3-diff-14-a3} ), one can do the similar computations like the one done in \cite{Hu:2023mgc} to find other components of generating function using the
result \eref{gen-IBP-3-diff-14}. Here we will not repeat the computation.

For our case with special values of $A,B_i, C_i$ given in \eref{gen-IBP-3-diff-11-2}, one can
find that
\be b_1= (D-(n+1)),~~~~~a_1={b_1\over 2}={(D-(n+1))\over 2}~~~~\label{gen-IBP-3-diff-14-a1}\ee
for \eref{gen-IBP-3-diff-12}, thus \eref{gen-IBP-3-diff-14} can be  simplified to
\be (\vec{\W\a}^{gen}(t,\W R))_{\emptyset}=e^{{-B_1+\sqrt{B_1^2-4 C_1 A}\over 2 A} t}~_1F_1({(D-(n+1))\over 2}; (D-(n+1));{-t\sqrt{B_1^2-4C_1 A }\over A})~~~~\label{gen-IBP-3-diff-14-a2}\ee

Results \eref{gen-IBP-3-diff-11} and \eref{gen-IBP-3-diff-12} are another main result in this paper since it tells us that
the generating function can be written as a compact generalized hypergeometric function.  Using these results,
the computation of reduction coefficients becomes a trivial task. Next, we present some examples.
\subsubsection{The tadpole}

For the tadpole without the choice $K_0=0$, we have
\bea F & = & -2M_0^2,\nn
 A & =& {-1\over 2M_0^2},~~~~~~B_0=(D-1){-1\over 2M_0^2},~~~~~B_1={2K_0\cdot \W R\over M_0^2},\nn
C_0&= &  {(D-1) 2K_0\cdot \W R\over 2M_0^2},
~~~~~~C_1=2\W R^2-{(2K_0\cdot \W R)^2\over 2M_0^2}~~~\label{tad-a-sol-1}\eea
and
\be \vec{\W\a}_{tad}^{gen}(t,\W R)=e^{ {(2K_0\cdot \W R-2\sqrt{M_0^2\W R^2})t}}~~_1F_1({(D-1)\over 2};D-1;4 t \sqrt{M_0^2 \W R^2})~~~\label{tad-a-sol-2}\ee
Comparing with \eref{Gen-tad-sol-1-9} and  \eref{Gen-tad-sol-1-10}, we get
following relation
\be e^{ {-2t\sqrt{M_0^2\W R^2}}}~~_1F_1({(D-1)\over 2};D-1;4 t \sqrt{M_0^2 \W R^2})=~~_0F_1(0; {D\over 2}; M_0^2 \W R^2 t^2 )~~~\label{tad-a-sol-3}\ee
In fact, the identity \eref{tad-a-sol-3} is the special case of the identity
\be e^{-x} ~~_1F_1(a;2a; 2x)= ~~_0F_1(\emptyset; a+{1\over 2}; {x^2\over 4})~~~\label{tad-a-sol-4}\ee
Using \eref{tad-a-sol-4}, \eref{gen-IBP-3-diff-14-a2} can be further simplified as
\bea (\vec{\W\a}^{gen}(t,\W R))_{\emptyset}=e^{{-B_1t \over 2 A} }~_0F_1(\emptyset;{(D+1-(n+1))\over 2}; {(B_1^2-4C_1 A) t^2 \over 16 A^2})~~~~\label{gen-IBP-3-diff-14-a3}\eea
where the square root in \eref{gen-IBP-3-diff-14-a2} disappear naturally using the form \eref{gen-IBP-3-diff-14-a3}.

\subsubsection{The massless bubble}

For this one, we just check the result \eref{gen-IBP-3-diff-14}. Using
\bea F= \left( \begin{array}{cc} 0 & K_1^2 \\ K_1^2 & 0 \end{array}\right)~,~~~F^{-1}= \left( \begin{array}{cc} 0 & {1\over K_1^2} \\ {1\over K_1^2} & 0 \end{array}\right)~~~\label{massless-bub}\eea
we have
\be A =  {2\over K^2},~~~ B_0= {2(D-2)\over K^2},~~B_1= {-4 K\cdot \W R\over K^2},~~C_0= {-2(D-2) K\cdot \W R\over K^2},~~C_1= 2\W R^2~~~\label{massless-bub-2}\ee
and
\be x=-2t\sqrt{ (K\cdot \W R)^2-K^2\W R^2},~~~b_1=(D-2),~~~a_1={ (D-2)
\over 2 }~~~\label{massless-bub-3}\ee
The generating function of the bubble to bubble is
\bea & & e^{(K\cdot \W R+\sqrt{ (K\cdot \W R)^2-K^2\W R^2})t} ~~_1F_1\left( { (D-2)
	\over 2 }; (D-2); -2t\sqrt{ (K\cdot \W R)^2-K^2\W R^2} \right)\nn
	& = & e^{(K\cdot \W R)t} ~~_0F_1\left(\emptyset; { (D-1)
		\over 2 };  {1\over 4}t^2((K\cdot \W R)^2-K^2\W R^2) \right)~~~\label{massless-bub-4}\eea
One can easily expand \eref{massless-bub-4} and check with known results:
\bea && 1+ (K\cdot \W R)t+ { (D (K\cdot \W R)^2-K^2 \W R^2) t^2\over 2(D-1)}+
{ ((D+2) (K\cdot \W R)^3-3 K^2 \W R^2 (K\cdot \W R) )t^3\over 6(D-1)}\nn
& & + { ((D+2)(D+4)(K\cdot \W R)^3-6 (D+2) K^2 \W R^2 (K\cdot \W R)^2+ 4 (K^2\W R^2)^2 )t^4\over 24 (D-1)(D+1)}+...\eea
%

\section{Solving ${\cal H}_{n+1;j}$}\label{H-matrix}

In this section, we will solve the matrix ${\cal H}_{n+1;j}$ defined in \eref{gen-1-5}, which is the reduction of scalar integrals
with the $j$-th propagator having power two while others, power one. The element of the matrix can be represented as $({\cal H}_{n+1;j})_{\WH{S}_1,\WH{S}_2 }$ where the list $S$ indicates the removing propagators from the $(n+1)$-gon.
Furthermore, by our iterative construction, only the first column of the matrix ${\cal H}_{n+1;j}$ is unknown, since all
other columns are  naturally embedding of the matrix ${\cal H}_{(n+1;\WH{S});j}$ where $(n+1;\WH{S})$ is the $m$-gon obtained
from $(n+1)$-gon by removing propagators from the non-empty list $S$.

Having recalled above facts, now we solve ${\cal H}_{n+1;j}$ using \eref{gen-T-3-3}, which expands to
\be \left( {\d\over \d (M_i^2)}+{\cal H}_{n+1;i}\right) \vec{\a}^{gen}=\sum_{r=0}^n (F^{-1}_{ir})\left(-\left(D-2-n-2K_r\cdot R+ R\cdot {\d\over \d R}\right) \vec{\a}^{gen}+B_r \right)~.~~~\label{gen-T-3-3b}\ee
If we put the tensor expansion \eref{gen-IBP-3-5-2} into \eref{gen-T-3-3b} and focus on terms independent of $R$, we will find
\be \left( {\d\over \d (M_i^2)}+{\cal H}_{n+1;i}\right) \vec{\a}_{00...0}=\sum_{r=0}^n (F^{-1}_{ir})\left(-\left(D-2-n\right) \vec{\a}_{00...0}+\vec{b}_{r;00...0} \right)~.~~~\label{gen-T-3-3c}\ee
Using the boundary condition, i.e., $\vec{\a}_{00...0}=\{1,0,0,...,0\}^T$, we arrive
\be {\cal H}_{n+1;i}\cdot \{1,0,0,...,0\}^T=\sum_{r=0}^n (F^{-1}_{ir})\left(-\left(D-2-n\right) \{1,0,0,...,0\}^T+\vec{b}_{r;00...0} \right)~.~~~\label{gen-T-3-3d}\ee
The left-hand side of \eref{gen-T-3-3d} picks the unknown first column of the $N\times N$ matrix ${\cal H}_{n+1;j}$,
thus $N$ equations\footnote{Since we assume the general masses, $N=2^{n+1}-1$.}
\eref{gen-T-3-3d} will determine ${\cal H}_{n+1;j}$ completely.

To have a better picture of \eref{gen-T-3-3d}, we need to understand $B_r$ better. Recalling the definition of \eref{gen-T-7-5}, each term $\left( {\cal D}_{n+1;j}\vec{\a}_{n+1;\WH r}^{gen}\right)$ is, in fact, the reduction of
(see \eref{gen-1-4})
\be
\int {d^D\ell\over i\pi^{D/2}}{e^{2\ell\cdot R}\over ((\ell-K_j)^2-M_j^2)^2 \prod_{t\neq j,r}  ((\ell-K_t)^2-M_t^2)}=\bm{J}_{n+1}\cdot \left( {\cal D}_{n+1;j}\vec{\a}_{n+1;\WH r}^{gen}\right)~.~~~~\label{gen-1-8}
\ee
To find the $R$-independent term in \eref{gen-1-8}, we set $R=0$ and the left-hand side becomes
\bea & & \int {d^D\ell\over i\pi^{D/2}}{1\over ((\ell-K_j)^2-M_j^2)^2 \prod_{t\neq j,r}  ((\ell-K_t)^2-M_t^2)} \nn
& = & {\d \over \d M_j^2}\left\{\int {d^D\ell\over i\pi^{D/2}}{1\over ((\ell-K_j)^2-M_j^2) \prod_{t\neq j,r}  ((\ell-K_t)^2-M_t^2)}\right\}\nn
& = & {\d \over \d M_j^2} (\bm{J}_{(n+1;\WH{r})}\cdot \vec{\a}_{\WH r})=\bm{J}_{(n+1;\WH{r})}\cdot {\cal H}_{(n+1;\WH{r});j}\cdot \vec{\a}_{\WH r}~.~~~~\label{gen-1-8bb}\eea
In \eref{gen-1-8bb}, $\bm{J}_{(n+1;\WH{r})}$  is the standard scalar basis according to our
convention \eref{gen-T-7-1} after removing the $r$-th propagator, thus we have the vector
$ \vec{\a}_{\WH r}=\{1,0,0,...,0\}^T$. From this argument, the $R$-independent term is nothing, but the first row of the matrix ${\cal H}_{(n+1;\WH{r});j}$ after the natural embedding
into the basis $\bm{J}_{n+1}$. Now the meaning of
$\vec{b}_{r;00...0}$ is clear: it is the sum of the reduction of scalar integrals with $r$-th propagator removed and $j$-th propagator ($j=0,1,...,n; j\neq r$) having
power two. Using the component form, we find
\be (\vec{b}_{r;00...0})_{\WH{S}}= \sum_{j=0,j\neq r}^n ({\cal H}_{n+1;j})_{\WH{S},\WH{r}}~.~~~\label{Br-br}\ee
Having this understanding, \eref{gen-T-3-3d} gives the solution of each component of the first column as
\be ({\cal H}_{n+1;i})_{\WH{S},\WH{\emptyset}}=-\left\{\left(D-2-n\right)\sum_{r=0}^n (F^{-1}_{ir}) \right\}\delta_{S,\emptyset}+ \sum_{r=0}^n (F^{-1}_{ir})\sum_{j=0,j\neq r}^n ({\cal H}_{n+1;j})_{\WH{S},\WH{r}}~,~~~\label{gen-T-H-1}\ee
or in vector form
\be {\cal H}_{n+1;i}|_{1C}=-\left\{\left(D-2-n\right)\sum_{r=0}^n (F^{-1}_{ir}) \right\}\{1,0,0,..,0\}^T+ \sum_{r=0}^n (F^{-1}_{ir})\sum_{j=0,j\neq r}^n ({\cal H}_{n+1;j})|_{(r+2)C}~~~~\label{gen-T-H-2}\ee
where $iC$ means the $i$-th column. Result \eref{gen-T-H-1} (or \eref{gen-T-H-2}) gives a very clear recursive construction of the matrix ${\cal H}_{n+1;i}$.
Now based on \eref{gen-T-H-1}, we can write down the recursive algorithm for all components of the matrix ${\cal H}_{n+1;i}$:
\begin{itemize}
	\item (1) Let us consider the arbitrary element $({\cal H}_{n+1;i})_{\WH{\cal R},\WH{\cal C}}$ where the removed lists ${\cal R},{\cal C}$ indicate
	the corresponding row and column. The first obvious result is when $i\in {\cal C}$, $({\cal H}_{n+1;i})_{\WH{\cal R},\WH{\cal C}}=0,~~\forall {\cal R}$. Now we will assume that $i\not\in {\cal C}$. The second obvious result is that
	when ${\cal C}\not\subseteq {\cal R}$, the $({\cal H}_{n+1;i})_{\WH{\cal R},\WH{\cal C}}=0$.
	
	\item (2) Now we define some notations for later convenience. First, we use ${\cal P}=\{0,1,...,n\}$ to represent the complete
	list of propagators. Given ${\cal C}$ we use $n_{{\cal C}}$ to denote the number of elements in the list and define the reduced modified Caylay matrix $F_{{\cal P}\setminus {\cal C}}$ by removing the
	corresponding rows and columns indicated by the list (see \eref{gen-T-2-2}).
	
	\item (3) When ${\cal R}= {\cal C}$, we have
	\be ({\cal H}_{n+1;i})_{\WH{\cal C},\WH{\cal C}}=-(D-1-(n+1-n_{{\cal C}}))\sum_{r\in {\cal P}\setminus {\cal C}}
	(F_{{\cal P}\setminus {\cal C}})^{-1}_{ir} ~.~~~\label{H-algo-1}\ee

	\item (4) When ${\cal C}\subset {\cal R}$, we have
	\be ({\cal H}_{n+1;i})_{\WH{\cal R},\WH{\cal C}}=\sum_{r\in {\cal P}\setminus {\cal C}}
	(F_{{\cal P}\setminus {\cal C}})^{-1}_{ir} \sum_{j\in {\cal P}\setminus \left({\cal C}\bigcup \{r\}\right)}({\cal H}_{n+1;j})_{\WH{\cal R},{\cal C}\bigcup \{r\}}~.~~~\label{H-algo-2}\ee

	\item (5) Equation \eref{H-algo-2} shows explicitly the recursive structure. With fixed  row list ${\cal R}$, the column list at the right-hand
	side is larger than the left-hand side by one element. Iteratively using  \eref{H-algo-2} will reach two possible
	terminations: either ${\cal R}= {\cal C}$ with the expression \eref{H-algo-1} and either ${\cal C}\not\subseteq {\cal R}$
	with zero contribution.
	
\end{itemize}

The above recursive algorithm for ${\cal H}_{n+1;i}$ is our third main result in this paper.  We have implemented the algorithm in a simple Mathematica file which introduced in \ref{H-Mathematica}, which can be downloaded from {\bf https://github.com/ybzhang-nxu/Hmatirx}. Now we demonstrate the use of \eref{gen-T-H-1} (or \eref{H-algo-1} and \eref{H-algo-2}) with several examples.

\subsection{Tadpole}

For the tadpole, the second term in \eref{gen-T-H-1} does not exist, thus we have immediately
\be {\cal H}_{1;0}=-(D-2) F_{00}^{-1}={(D-2)\over 2M_0^2}~~~\label{tad-1-5}\ee
If we use \eref{H-algo-1} and \eref{H-algo-2}, the only case is ${\cal R}= {\cal C}=\emptyset$, which reduces to
\eref{tad-1-5} again.

\subsection{Bubble}

For the bubble, the second term in \eref{gen-T-H-1} does contribute. Using  \eref{tad-1-5}, we have
\be \vec{b}_{0;00...0}=\left( \begin{array}{c} 0 \\ {(D-2)\over 2M_1^2} \\ 0 \end{array} \right)~,~~~\vec{b}_{1;00...0}=\left( \begin{array}{c} 0 \\ 0\\ {(D-2)\over 2M_0^2}  \end{array} \right)~,~~~\label{Bub-new-1}\ee
according to the basis \eref{bub-1-1}. Using \eref{gen-T-H-2} and the modified Caylay matrix $F$ given in \eref{bub-2-5} we get the first column as
\bea ({\cal H}_{2;0})|_{1C} =  {1\over  \Delta_{bub}}
\left( \begin{array}{c} {(D-3) (K_1^2+M_1^2-M_0^2)} \\ {-(D-2)} \\ -(K_1^2-M_0^2-M_1^2){(D-2)\over 2M_0^2}  \end{array} \right)~,~~
 ({\cal H}_{2;1})|_{1C} ={1\over  \Delta_{bub}}
\left( \begin{array}{c} {(D-3) (K_1^2+M_0^2-M_1^2)} \\ -(K_1^2-M_0^2-M_1^2){(D-2)\over 2M_1^2} \\ -(D-2) \end{array} \right)~.~~~\label{Bub-new-4}\eea
where $\Delta_{bub}={\rm det}(F)$. The other columns are natural embedding of \eref{tad-1-5}, i.e.,
\bea {\cal H}_{2;0}=\left( \begin{array}{ccc} * & 0 & 0  \\ * & 0 & 0\\
	* & 0 &  {(D-2)\over 2 M_0^2}
\end{array}\right),~~~{\cal H}_{2;1}=\left( \begin{array}{ccc} * & 0 & 0  \\ * & {(D-2)\over 2 M_1^2}  & 0\\
	* & 0 &  0
\end{array}\right)~~\label{bub-1-4}\eea
%

\subsection{Triangle}

To simplify the notation, let us use $ \Delta_{I}$ to represent the determinant of the sub-matrix,
which is obtained from  the modified Caylay matrix $F$ defined in \eref{gen-T-2-2} by extracting the rows and columns corresponding to the index set $I$. Thus the first column of
$ \left(\mathcal{H}_{3 ; 0}\right) $ can be expressed as:
\be
\left.\left(\mathcal{H}_{3 ; 0}\right)\right|_{1 C}=	\left(
	\begin{array}{l}
		-\frac{(D-4) \left(f_{01} \left(f_{12}+2 M_2^2\right)+f_{02} \left(f_{12}+2 M_1^2\right)-f_{12}^2+4 M_1^2 M_2^2\right)}{\Delta _{tri}} \\-\frac{2 (D-3) \left(f_{12}+M_1^2+M_2^2\right) \left(f_{12}^2-4 M_1^2 M_2^2\right)}{\Delta _{23} \Delta _{tri}} \\\frac{2 (D-3) \left(f_{02}+M_0^2+M_2^2\right) \left(2 M_2^2 f_{01}+f_{02} f_{12}\right)}{\Delta _{13} \Delta _{tri}} \\\frac{2 (D-3) \left(f_{01}+M_0^2+M_1^2\right) \left(f_{01} f_{12}+2 M_1^2 f_{02}\right)}{\Delta _{12} \Delta _{tri}} \\\frac{(D-2) \left(\Delta _{13} \left(f_{12}+2 M_2^2\right) \left(f_{12}^2-4 M_1^2 M_2^2\right)-\Delta _{23} \left(f_{02}+2 M_2^2\right) \left(2 M_2^2 f_{01}+f_{02} f_{12}\right)\right)}{2 M_2^2 \Delta _{13} \Delta _{23} \Delta _{tri}} \\\frac{(D-2) \left(\Delta _{12} \left(f_{12}+2 M_1^2\right) \left(f_{12}^2-4 M_1^2 M_2^2\right)-\Delta _{23} \left(f_{01}+2 M_1^2\right) \left(f_{01} f_{12}+2 M_1^2 f_{02}\right)\right)}{2 M_1^2 \Delta _{12} \Delta _{23} \Delta _{tri}} \\\frac{(2-D) \left(\Delta _{13} \left(f_{01}+2 M_0^2\right) \left(f_{01} f_{12}+2 M_1^2 f_{02}\right)+\Delta _{12} \left(f_{02}+2 M_0^2\right) \left(2 M_2^2 f_{01}+f_{02} f_{12}\right)\right)}{2 M_0^2 \Delta _{12} \Delta _{13} \Delta _{tri}} \\\end{array} \right)
\ee
Similarly, the first columns of other matrices are
\be
\left.\left(\mathcal{H}_{3 ; 1}\right)\right|_{1 C}	\left(\begin{array}{l}
		-\frac{(D-4) \left(f_{01} \left(f_{02}+2 M_2^2\right)+f_{02} f_{12}-f_{02}^2+2 M_0^2 \left(f_{12}+2 M_2^2\right)\right)}{\Delta _{tri}} \\\frac{2 (D-3) \left(f_{12}+M_1^2+M_2^2\right) \left(2 M_2^2 f_{01}+f_{02} f_{12}\right)}{\Delta _{23} \Delta _{tri}} \\-\frac{2 (D-3) \left(f_{02}+M_0^2+M_2^2\right) \left(f_{02}^2-4 M_0^2 M_2^2\right)}{\Delta _{13} \Delta _{tri}} \\\frac{2 (D-3) \left(f_{01}+M_0^2+M_1^2\right) \left(f_{01} f_{02}+2 M_0^2 f_{12}\right)}{\Delta _{12} \Delta _{tri}} \\\frac{(D-2) \left(\Delta _{23} \left(f_{02}+2 M_2^2\right) \left(f_{02}^2-4 M_0^2 M_2^2\right)-\Delta _{13} \left(f_{12}+2 M_2^2\right) \left(2 M_2^2 f_{01}+f_{02} f_{12}\right)\right)}{2 M_2^2 \Delta _{13} \Delta _{23} \Delta _{tri}} \\\frac{(2-D) \left(\Delta _{23} \left(f_{01}+2 M_1^2\right) \left(f_{01} f_{02}+2 M_0^2 f_{12}\right)+\Delta _{12} \left(f_{12}+2 M_1^2\right) \left(2 M_2^2 f_{01}+f_{02} f_{12}\right)\right)}{2 M_1^2 \Delta _{12} \Delta _{23} \Delta _{tri}} \\\frac{(D-2) \left(\Delta _{12} \left(f_{02}+2 M_0^2\right) \left(f_{02}^2-4 M_0^2 M_2^2\right)-\Delta _{13} \left(f_{01}+2 M_0^2\right) \left(f_{01} f_{02}+2 M_0^2 f_{12}\right)\right)}{2 M_0^2 \Delta _{12} \Delta _{13} \Delta _{tri}} \\\end{array} \right)
\ee
and
\be  \left.\left(\mathcal{H}_{3 ; 2}\right)\right|_{1 C}\left(\begin{array}{l}
	\frac{(D-4) \left(-f_{01} \left(f_{02}+f_{12}\right)+f_{01}^2-2 M_1^2 \left(f_{02}+2 M_0^2\right)-2 M_0^2 f_{12}\right)}{\Delta _{tri}} \\\frac{2 (D-3) \left(f_{12}+M_1^2+M_2^2\right) \left(f_{01} f_{12}+2 M_1^2 f_{02}\right)}{\Delta _{23} \Delta _{tri}} \\\frac{2 (D-3) \left(f_{02}+M_0^2+M_2^2\right) \left(f_{01} f_{02}+2 M_0^2 f_{12}\right)}{\Delta _{13} \Delta _{tri}} \\-\frac{2 (D-3) \left(f_{01}+M_0^2+M_1^2\right) \left(f_{01}^2-4 M_0^2 M_1^2\right)}{\Delta _{12} \Delta _{tri}} \\\frac{(2-D) \left(\Delta _{13} \left(f_{12}+2 M_2^2\right) \left(f_{01} f_{12}+2 M_1^2 f_{02}\right)+\Delta _{23} \left(f_{02}+2 M_2^2\right) \left(f_{01} f_{02}+2 M_0^2 f_{12}\right)\right)}{2 M_2^2 \Delta _{13} \Delta _{23} \Delta _{tri}} \\\frac{(D-2) \left(\Delta _{23} \left(f_{01}+2 M_1^2\right) \left(f_{01}^2-4 M_0^2 M_1^2\right)-\Delta _{12} \left(f_{12}+2 M_1^2\right) \left(f_{01} f_{12}+2 M_1^2 f_{02}\right)\right)}{2 M_1^2 \Delta _{12} \Delta _{23} \Delta _{tri}} \\\frac{(D-2) \left(\Delta _{13} \left(f_{01}+2 M_0^2\right) \left(f_{01}^2-4 M_0^2 M_1^2\right)-\Delta _{12} \left(f_{02}+2 M_0^2\right) \left(f_{01} f_{02}+2 M_0^2 f_{12}\right)\right)}{2 M_0^2 \Delta _{12} \Delta _{13} \Delta _{tri}} \\\end{array} \right)\ee
Starting from this subsection, we will not write down the remaining columns of $\mathcal{H}_{n ; i}$, which are natural extensions of
known results (for example, the \eref{bub-1-4} for $n=3$) according to the basis ordering defined in \eref{gen-T-7-1}.

\subsection{The box}

We have used  our formula to compute the matrices  $\mathcal{H}_{n ; i}$ up to pentagon $n=5$ and check numerically with FIRE \cite{Smirnov:2019qkx} up to
some $i$ of the box. In this part, we will present  only the result $\mathcal{H}_{4; 0}$ to demonstrate the iterative structure of denominators and the increasing complexity of numerators when going to lower and lower gon:
\bean
	&\left.\left(\mathcal{H}_{4 ; 0}\right)\right|_{ 0,1}=\frac{1}{\Delta _{\text{box}}}\\
	&& \Big((D-5) (f_{01} f_{12} f_{23}+2 M_3^2 (f_{12} (f_{01}-f_{12})+2 M_2^2 (f_{01}+2 M_1^2))+f_{01} f_{13} f_{23}+2 M_2^2 f_{01} f_{13}- \\
	& &f_{01} f_{23}^2+f_{02} (f_{12} (f_{13}+2 M_3^2)+f_{13} f_{23}-f_{13}^2+2 M_1^2 (f_{23}+2 M_3^2))+f_{03} (f_{12} (f_{13}+ \\
	& &f_{23})-f_{12}^2+2 M_2^2 (f_{13}+2 M_1^2)+2 M_1^2 f_{23})-2 f_{12} f_{13} f_{23}-2 M_2^2 f_{13}^2-2 M_1^2 f_{23}^2)\Big)
\eean
\bean
	& & \left.\left(\mathcal{H}_{4 ; 0}\right)\right|_{ 1,1}=\frac{1}{\Delta _{234} \Delta _{\text{box}}}\\
	&& \Big(2 (D-4) (-2 f_{12} (f_{13}+f_{23}+2 M_3^2)+f_{12}^2-2 f_{13} (f_{23}+2 M_2^2)+f_{13}^2-4 M_1^2 f_{23}+f_{23}^2- \\
	&&4 M_1^2 M_2^2-4 (M_1^2+M_2^2) M_3^2) (f_{12} f_{13} f_{23}+M_3^2 f_{12}^2+M_2^2 (f_{13}^2-4 M_1^2 M_3^2)+M_1^2 f_{23}^2)\Big)
\eean
\bean
	& & \left.\left(\mathcal{H}_{4 ; 0}\right)\right|_{ 2,1}=\frac{1}{\Delta _{134} \Delta _{\text{box}}}\\
	&& \Big(-((D-4) (-2 f_{02} (f_{03}+f_{23}+2 M_3^2)+f_{02}^2-2 f_{03} (f_{23}+2 M_2^2)+f_{03}^2-4 M_0^2 f_{23}+f_{23}^2- \\
	&&4 M_0^2 M_2^2-4 (M_0^2+M_2^2) M_3^2) (2 M_3^2 (2 M_2^2 f_{01}+f_{02} f_{12})+f_{23} (f_{02} f_{13}-f_{01} f_{23})+ \\
	& &f_{03} (f_{12} f_{23}+2 M_2^2 f_{13})))\Big)
\eean
\begin{align*}
	\left.\left(\mathcal{H}_{4 ; 0}\right)\right|_{ 3,1}=&\frac{1}{\Delta _{124} \Delta _{\text{box}}}\\
	& \Big(-((D-4) (-2 f_{01} (f_{03}+f_{13}+2 M_3^2)+f_{01}^2-2 f_{03} (f_{13}+2 M_1^2)+f_{03}^2-4 M_0^2 f_{13}+f_{13}^2- \\
	&4 M_0^2 M_1^2-4 (M_0^2+M_1^2) M_3^2) (2 M_3^2 (f_{01} f_{12}+2 M_1^2 f_{02})+f_{13} (f_{01} f_{23}-f_{02} f_{13})+ \\
	&f_{03} (f_{12} f_{13}+2 M_1^2 f_{23})))\Big)
\end{align*}
\begin{align*}
	\left.\left(\mathcal{H}_{4 ; 0}\right)\right|_{ 4,1}=&\frac{1}{\Delta _{123} \Delta _{\text{box}}}\\
	& \Big(-((D-4) (-2 f_{01} (f_{02}+f_{12}+2 M_2^2)+f_{01}^2-2 f_{02} (f_{12}+2 M_1^2)+f_{02}^2-4 M_0^2 f_{12}+f_{12}^2- \\
	&4 M_0^2 M_1^2-4 (M_0^2+M_1^2) M_2^2) (f_{01} (f_{12} f_{23}+2 M_2^2 f_{13})+f_{02} (f_{12} f_{13}+2 M_1^2 f_{23})- \\
	&f_{03} (f_{12}^2-4 M_1^2 M_2^2)))\Big)
\end{align*}
\begin{align*}
	\left.\left(\mathcal{H}_{4 ; 0}\right)\right|_{ 5,1}=&\frac{1}{\Delta _{134} \Delta _{234} \Delta _{34} \Delta _{\text{box}}}\\
	& \Big(-2 (D-3) (f_{23}+M_2^2+M_3^2) (\Delta _{234} (f_{02} f_{23}+2 M_3^2 (f_{02}+2 M_2^2)+f_{03} f_{23}+2 M_2^2 f_{03}- \\
	&f_{23}^2) (-f_{01} (f_{23}^2-4 M_2^2 M_3^2)+f_{02} f_{13} f_{23}+2 M_2^2 f_{03} f_{13})+f_{12} (\Delta _{234} (2 M_3^2 f_{02}+ \\
	&f_{03} f_{23}) (f_{02} f_{23}+2 M_3^2 (f_{02}+2 M_2^2)+f_{03} f_{23}+2 M_2^2 f_{03}-f_{23}^2)-2 \Delta _{134} (f_{13}^2 (3 M_2^2 f_{23}+ \\
	&f_{23}^2+2 M_2^2 M_3^2)-f_{13} (f_{23}^3-4 M_2^2 M_3^2 f_{23})+M_1^2 (f_{23}+2 M_3^2) (f_{23}^2-4 M_2^2 M_3^2)))- \\
	&2 f_{12}^2 \Delta _{134} (M_3^2 (f_{13} (3 f_{23}+2 M_2^2)-f_{23}^2)+f_{13} f_{23}^2+4 M_2^2 M_3^4)-2 M_3^2 f_{12}^3 \Delta _{134} (f_{23}+ \\
	&2 M_3^2)-2 \Delta _{134} (f_{13} (f_{23}+2 M_2^2)-f_{23}^2+4 M_2^2 M_3^2) (M_2^2 (f_{13}^2-4 M_1^2 M_3^2)+M_1^2 f_{23}^2))\Big)
\end{align*}
\begin{align*}
	\left.\left(\mathcal{H}_{4 ; 0}\right)\right|_{ 6,1}=&\frac{1}{\Delta _{124} \Delta _{234} \Delta _{24} \Delta _{\text{box}}}\\
	& \Big(-2 (D-3) (f_{13}+M_1^2+M_3^2) (\Delta _{234} (f_{01} f_{13}+2 M_3^2 (f_{01}+2 M_1^2)+f_{03} f_{13}+2 M_1^2 f_{03}- \\
	&f_{13}^2) (f_{23} (f_{01} f_{13}+2 M_1^2 f_{03})-f_{02} (f_{13}^2-4 M_1^2 M_3^2))+f_{12} (\Delta _{234} (2 M_3^2 f_{01}+ \\
	&f_{03} f_{13}) (f_{01} f_{13}+2 M_3^2 (f_{01}+2 M_1^2)+f_{03} f_{13}+2 M_1^2 f_{03}-f_{13}^2)+2 \Delta _{124} (- \\
	&2 M_3^2 (M_1^2 f_{23} (2 f_{13}+f_{23})+M_2^2 f_{13} (f_{13}-2 M_1^2))+f_{13} f_{23} (-f_{13} f_{23}+f_{13}^2- \\
	&3 M_1^2 f_{23})-M_2^2 f_{13}^3+8 M_1^2 M_2^2 M_3^4))-2 f_{12}^2 \Delta _{124} (f_{13}^2 (f_{23}-M_3^2)+3 M_3^2 f_{13} f_{23}+ \\
	&2 M_1^2 M_3^2 (f_{23}+2 M_3^2))-2 M_3^2 f_{12}^3 \Delta _{124} (f_{13}+2 M_3^2)+2 \Delta _{124} (-f_{13} f_{23}+ \\
	&f_{13}^2-2 M_1^2 (f_{23}+2 M_3^2)) (M_2^2 (f_{13}^2-4 M_1^2 M_3^2)+M_1^2 f_{23}^2))\Big)
\end{align*}
\begin{align*}
	\left.\left(\mathcal{H}_{4 ; 0}\right)\right|_{ 7,1}=&\frac{1}{\Delta _{123} \Delta _{23} \Delta _{234} \Delta _{\text{box}}}\\
	& \Big(-2 (D-3) (f_{12}+M_1^2+M_2^2) (-\Delta _{234} (f_{01} (f_{12}+2 M_2^2)+f_{02} (f_{12}+2 M_1^2)-f_{12}^2+4 M_1^2 M_2^2) (- \\
	&f_{01} (f_{12} f_{23}+2 M_2^2 f_{13})-f_{02} (f_{12} f_{13}+2 M_1^2 f_{23})+f_{03} (f_{12}^2-4 M_1^2 M_2^2))-2 \Delta _{123} (f_{12} (f_{13}+ \\
	&f_{23})-f_{12}^2+2 M_2^2 (f_{13}+2 M_1^2)+2 M_1^2 f_{23}) (f_{12} f_{13} f_{23}+M_3^2 f_{12}^2+M_2^2 (f_{13}^2- \\
	&4 M_1^2 M_3^2)+M_1^2 f_{23}^2))\Big)
\end{align*}
\begin{align*}
	\left.\left(\mathcal{H}_{4 ; 0}\right)\right|_{ 8,1}=&\frac{1}{\Delta _{124} \Delta _{134} \Delta _{14} \Delta _{\text{box}}}\\
	& \Big(2 (D-3) (f_{03}+M_0^2+M_3^2) (-(\Delta _{124} (f_{02} (f_{03}+2 M_3^2)+f_{03} f_{23}-f_{03}^2+2 M_0^2 (f_{23}+ \\
	&2 M_3^2)) (2 M_3^2 (2 M_2^2 f_{01}+f_{02} f_{12})+f_{23} (f_{02} f_{13}-f_{01} f_{23})+f_{03} (f_{12} f_{23}+2 M_2^2 f_{13})))- \\
	&\Delta _{134} (f_{01} (f_{03}+2 M_3^2)+f_{03} f_{13}-f_{03}^2+2 M_0^2 (f_{13}+2 M_3^2)) (2 M_3^2 (f_{01} f_{12}+ \\
	&2 M_1^2 f_{02})+f_{13} (f_{01} f_{23}-f_{02} f_{13})+f_{03} (f_{12} f_{13}+2 M_1^2 f_{23})))\Big)
\end{align*}
\begin{align*}
	\left.\left(\mathcal{H}_{4 ; 0}\right)\right|_{ 9,1}=&\frac{1}{\Delta _{123} \Delta _{13} \Delta _{134} \Delta _{\text{box}}}\\
	& \Big(2 (D-3) (f_{02}+M_0^2+M_2^2) (f_{02}^2 (\Delta _{123} (-2 M_3^2 (f_{12} f_{23}-2 M_2^2 f_{01})-(f_{23}^2 (f_{01}+ \\
	&f_{13}))+f_{03} (f_{12} (f_{23}-2 M_3^2)-f_{13} (f_{23}-2 M_2^2)))-\Delta _{134} (f_{01} (f_{12} (f_{13}-f_{23})- \\
	&2 M_2^2 f_{13}+2 M_1^2 f_{23})+f_{03} (f_{12}^2-4 M_1^2 M_2^2)+f_{12} (f_{12} f_{13}+2 M_1^2 f_{23})))+f_{02} (f_{03} (\Delta _{123} (f_{23} (f_{23} (f_{01}- \\
	&f_{12})-4 M_2^2 f_{13})-4 M_2^2 M_3^2 (f_{01}+f_{12}))+\Delta _{134} (f_{01}+f_{12}) (f_{12}^2-4 M_1^2 M_2^2))+ \\
	&\Delta _{123} (f_{01} (f_{23}^3-4 M_2^2 M_3^2 f_{23})-2 M_0^2 (f_{23}+2 M_2^2) (2 M_3^2 f_{12}+f_{13} f_{23}))- \\
	&\Delta _{134} (f_{01}^2 (f_{12} f_{23}+2 M_2^2 f_{13})+f_{01} (4 M_2^2 (f_{12} f_{13}+M_1^2 f_{23})+f_{12}^2 f_{23})+ \\
	&2 M_0^2 (f_{12}+2 M_2^2) (f_{12} f_{13}+2 M_1^2 f_{23}))-(f_{03}^2 \Delta _{123} (f_{12} f_{23}+2 M_2^2 f_{13})))- \\
	&2 \Delta _{123} (M_2^2 (f_{03}+2 M_0^2)+M_0^2 f_{23}) (f_{03} (f_{12} f_{23}+2 M_2^2 f_{13})-f_{01} (f_{23}^2- \\
	&4 M_2^2 M_3^2))+2 \Delta _{134} (M_2^2 (f_{01}+2 M_0^2)+M_0^2 f_{12}) (f_{03} (f_{12}^2-4 M_1^2 M_2^2)-f_{01} (f_{12} f_{23}+ \\
	&2 M_2^2 f_{13}))+f_{02}^3 (\Delta _{123} (2 M_3^2 f_{12}+f_{13} f_{23})+\Delta _{134} (f_{12} f_{13}+2 M_1^2 f_{23})))\Big)
\end{align*}

\begin{align*}
	\left.\left(\mathcal{H}_{4 ; 0}\right)\right|_{ 10,1}=&\frac{1}{\Delta _{12} \Delta _{123} \Delta _{124} \Delta _{\text{box}}}\\
	& \Big(2 (D-3) (f_{01}+M_0^2+M_1^2) (f_{01}^2 (\Delta _{123} (-2 M_3^2 (f_{12} f_{13}-2 M_1^2 f_{02})-(f_{13}^2 (f_{02}+ \\
	&f_{23}))+f_{03} (f_{12} (f_{13}-2 M_3^2)-f_{23} (f_{13}-2 M_1^2)))-\Delta _{124} (2 M_2^2 (f_{13} (f_{02}+f_{12})- \\
	&2 M_1^2 f_{03})-f_{02} f_{12} f_{13}+f_{02} f_{12} f_{23}-2 M_1^2 f_{02} f_{23}+f_{03} f_{12}^2+f_{12}^2 f_{23}))+ \\
	&f_{01} (f_{03} (\Delta _{123} (f_{13} (f_{13} (f_{02}-f_{12})-4 M_1^2 f_{23})-4 M_1^2 M_3^2 (f_{02}+f_{12}))+\Delta _{124} (f_{02}+ \\
	&f_{12}) (f_{12}^2-4 M_1^2 M_2^2))+\Delta _{123} (f_{02} (f_{13}^3-4 M_1^2 M_3^2 f_{13})-2 M_0^2 (f_{13}+2 M_1^2) (2 M_3^2 f_{12}+ \\
	&f_{13} f_{23}))-\Delta _{124} (f_{02}^2 (f_{12} f_{13}+2 M_1^2 f_{23})+f_{02} (f_{12}^2 f_{13}+4 M_1^2 f_{12} f_{23}+ \\
	&4 M_1^2 M_2^2 f_{13})+2 M_0^2 (f_{12}+2 M_1^2) (f_{12} f_{23}+2 M_2^2 f_{13}))-(f_{03}^2 \Delta _{123} (f_{12} f_{13}+ \\
	&2 M_1^2 f_{23})))+f_{01}^3 (f_{12} (2 M_3^2 \Delta _{123}+\Delta _{124} f_{23})+f_{13} (\Delta _{123} f_{23}+2 M_2^2 \Delta _{124}))- \\
	&2 \Delta _{123} (M_1^2 (f_{03}+2 M_0^2)+M_0^2 f_{13}) (f_{03} (f_{12} f_{13}+2 M_1^2 f_{23})-f_{02} (f_{13}^2- \\
	&4 M_1^2 M_3^2))+2 \Delta _{124} (M_1^2 (f_{02}+2 M_0^2)+M_0^2 f_{12}) (f_{03} (f_{12}^2-4 M_1^2 M_2^2)-f_{02} (f_{12} f_{13}+ \\
	&2 M_1^2 f_{23})))\Big)
\end{align*}

\begin{align*}
	\left.\left(\mathcal{H}_{4 ; 0}\right)\right|_{ 11,1}=&\frac{1}{2 M_3^2 \Delta _{124} \Delta _{134} \Delta _{14} \Delta _{234} \Delta _{24} \Delta _{34} \Delta _{\text{box}}}\\
	& \Big((2-D) (-\Delta _{124} \Delta _{234} \Delta _{24} (2 M_3^2 (2 M_2^2 f_{01}+f_{02} f_{12})+f_{23} (f_{02} f_{13}- \\
	&f_{01} f_{23})+f_{03} (f_{12} f_{23}+2 M_2^2 f_{13})) (\Delta _{14} (f_{23}+2 M_3^2) (f_{02} (f_{23}+2 M_3^2)+ \\
	&f_{03} (f_{23}+2 M_2^2)-f_{23}^2+4 M_2^2 M_3^2)+\Delta _{34} (f_{03}+2 M_3^2) (f_{02} (f_{03}+2 M_3^2)+f_{03} f_{23}- \\
	&f_{03}^2+2 M_0^2 (f_{23}+2 M_3^2)))-(\Delta _{134} \Delta _{234} \Delta _{34} (2 M_3^2 (f_{01} f_{12}+2 M_1^2 f_{02})+ \\
	&f_{13} (f_{01} f_{23}-f_{02} f_{13})+f_{03} (f_{12} f_{13}+2 M_1^2 f_{23})) (\Delta _{14} (f_{13}+2 M_3^2) (f_{01} (f_{13}+ \\
	&2 M_3^2)+f_{03} (f_{13}+2 M_1^2)-f_{13}^2+4 M_1^2 M_3^2)+\Delta _{24} (f_{03}+2 M_3^2) (f_{01} (f_{03}+2 M_3^2)+ \\
	&f_{03} f_{13}-f_{03}^2+2 M_0^2 (f_{13}+2 M_3^2))))+2 \Delta _{124} \Delta _{134} \Delta _{14} (f_{12} f_{13} f_{23}+ \\
	&M_3^2 f_{12}^2+M_2^2 (f_{13}^2-4 M_1^2 M_3^2)+M_1^2 f_{23}^2) (\Delta _{24} (f_{23}+2 M_3^2) (f_{12} (f_{23}+2 M_3^2)+ \\
	&f_{13} (f_{23}+2 M_2^2)-f_{23}^2+4 M_2^2 M_3^2)+\Delta _{34} (f_{13}+2 M_3^2) (f_{12} (f_{13}+2 M_3^2)+f_{13} f_{23}- \\
	&f_{13}^2+2 M_1^2 (f_{23}+2 M_3^2))))\Big)
\end{align*}

\begin{align*}
	\left.\left(\mathcal{H}_{4 ; 0}\right)\right|_{ 12,1}=&\frac{1}{2 M_2^2 \Delta _{123} \Delta _{13} \Delta _{134} \Delta _{23} \Delta _{234} \Delta _{34} \Delta _{\text{box}}}\\
	& \Big((2-D) (-\Delta _{123} \Delta _{23} \Delta _{234} (2 M_3^2 (2 M_2^2 f_{01}+f_{02} f_{12})+f_{23} (f_{02} f_{13}- \\
	&f_{01} f_{23})+f_{03} (f_{12} f_{23}+2 M_2^2 f_{13})) (\Delta _{13} (f_{23}+2 M_2^2) (f_{02} (f_{23}+2 M_3^2)+ \\
	&f_{03} (f_{23}+2 M_2^2)-f_{23}^2+4 M_2^2 M_3^2)+\Delta _{34} (f_{02}+2 M_2^2) (f_{02} (f_{03}+f_{23})-f_{02}^2+ \\
	&2 M_2^2 (f_{03}+2 M_0^2)+2 M_0^2 f_{23}))+\Delta _{134} \Delta _{234} \Delta _{34} (-f_{01} (f_{12} f_{23}+2 M_2^2 f_{13})- \\
	&f_{02} (f_{12} f_{13}+2 M_1^2 f_{23})+f_{03} (f_{12}^2-4 M_1^2 M_2^2)) (\Delta _{13} (f_{12}+2 M_2^2) (f_{01} (f_{12}+ \\
	&2 M_2^2)+f_{02} (f_{12}+2 M_1^2)-f_{12}^2+4 M_1^2 M_2^2)+\Delta _{23} (f_{02}+2 M_2^2) (f_{01} (f_{02}+2 M_2^2)+ \\
	&f_{02} f_{12}-f_{02}^2+2 M_0^2 (f_{12}+2 M_2^2)))+2 \Delta _{123} \Delta _{13} \Delta _{134} (f_{12} f_{13} f_{23}+ \\
	&M_3^2 f_{12}^2+M_2^2 (f_{13}^2-4 M_1^2 M_3^2)+M_1^2 f_{23}^2) (\Delta _{34} (f_{12}+2 M_2^2) (f_{12} (f_{13}+f_{23})- \\
	&f_{12}^2+2 M_2^2 (f_{13}+2 M_1^2)+2 M_1^2 f_{23})+\Delta _{23} (f_{23}+2 M_2^2) (f_{12} (f_{23}+2 M_3^2)+f_{13} (f_{23}+ \\
	&2 M_2^2)-f_{23}^2+4 M_2^2 M_3^2)))\Big)
\end{align*}

\begin{align*}
	\left.\left(\mathcal{H}_{4 ; 0}\right)\right|_{ 13,1}=&\frac{1}{2 M_1^2 \Delta _{12} \Delta _{123} \Delta _{124} \Delta _{23} \Delta _{234} \Delta _{24} \Delta _{\text{box}}}\\
	& \Big((2-D) (-\Delta _{123} \Delta _{23} \Delta _{234} (2 M_3^2 (f_{01} f_{12}+2 M_1^2 f_{02})+f_{13} (f_{01} f_{23}- \\
	&f_{02} f_{13})+f_{03} (f_{12} f_{13}+2 M_1^2 f_{23})) (\Delta _{12} (f_{13}+2 M_1^2) (f_{01} (f_{13}+2 M_3^2)+ \\
	&f_{03} (f_{13}+2 M_1^2)-f_{13}^2+4 M_1^2 M_3^2)+\Delta _{24} (f_{01}+2 M_1^2) (f_{01} (f_{03}+f_{13})-f_{01}^2+ \\
	&2 M_1^2 (f_{03}+2 M_0^2)+2 M_0^2 f_{13}))+\Delta _{124} \Delta _{234} \Delta _{24} (\Delta _{23} (f_{01}+2 M_1^2) (f_{01} (f_{02}+ \\
	&f_{12})-f_{01}^2+2 M_1^2 (f_{02}+2 M_0^2)+2 M_0^2 f_{12})+\Delta _{12} (f_{12}+2 M_1^2) (f_{01} (f_{12}+2 M_2^2)+ \\
	&f_{02} (f_{12}+2 M_1^2)-f_{12}^2+4 M_1^2 M_2^2)) (-f_{01} (f_{12} f_{23}+2 M_2^2 f_{13})-f_{02} (f_{12} f_{13}+ \\
	&2 M_1^2 f_{23})+f_{03} (f_{12}^2-4 M_1^2 M_2^2))+2 \Delta _{12} \Delta _{123} \Delta _{124} (f_{12} f_{13} f_{23}+ \\
	&M_3^2 f_{12}^2+M_2^2 (f_{13}^2-4 M_1^2 M_3^2)+M_1^2 f_{23}^2) (\Delta _{24} (f_{12}+2 M_1^2) (f_{12} (f_{13}+f_{23})- \\
	&f_{12}^2+2 M_2^2 (f_{13}+2 M_1^2)+2 M_1^2 f_{23})+\Delta _{23} (f_{13}+2 M_1^2) (f_{12} (f_{13}+2 M_3^2)+f_{13} f_{23}- \\
	&f_{13}^2+2 M_1^2 (f_{23}+2 M_3^2))))\Big)
\end{align*}

\begin{align*}
	\left.\left(\mathcal{H}_{4 ; 0}\right)\right|_{ 14,1}=&\frac{1}{2 M_0^2 \Delta _{12} \Delta _{123} \Delta _{124} \Delta _{13} \Delta _{134} \Delta _{14} \Delta _{\text{box}}}\\
	& \Big((2-D) (-\Delta _{12} \Delta _{123} \Delta _{124} (2 M_3^2 (2 M_2^2 f_{01}+f_{02} f_{12})+f_{23} (f_{02} f_{13}- \\
	&f_{01} f_{23})+f_{03} (f_{12} f_{23}+2 M_2^2 f_{13})) (\Delta _{13} (f_{03}+2 M_0^2) (f_{02} (f_{03}+2 M_3^2)+ \\
	&f_{03} f_{23}-f_{03}^2+2 M_0^2 (f_{23}+2 M_3^2))+\Delta _{14} (f_{02}+2 M_0^2) (f_{02} (f_{03}+f_{23})-f_{02}^2+ \\
	&2 M_2^2 (f_{03}+2 M_0^2)+2 M_0^2 f_{23}))-\Delta _{123} \Delta _{13} \Delta _{134} (2 M_3^2 (f_{01} f_{12}+2 M_1^2 f_{02})+ \\
	&f_{13} (f_{01} f_{23}-f_{02} f_{13})+f_{03} (f_{12} f_{13}+2 M_1^2 f_{23})) (\Delta _{12} (f_{03}+2 M_0^2) (f_{01} (f_{03}+ \\
	&2 M_3^2)+f_{03} f_{13}-f_{03}^2+2 M_0^2 (f_{13}+2 M_3^2))+\Delta _{14} (f_{01}+2 M_0^2) (f_{01} (f_{03}+f_{13})- \\
	&f_{01}^2+2 M_1^2 (f_{03}+2 M_0^2)+2 M_0^2 f_{13}))+\Delta _{124} \Delta _{134} \Delta _{14} (\Delta _{13} (f_{01}+ \\
	&2 M_0^2) (f_{01} (f_{02}+f_{12})-f_{01}^2+2 M_1^2 (f_{02}+2 M_0^2)+2 M_0^2 f_{12})+\Delta _{12} (f_{02}+2 M_0^2) (f_{01} (f_{02}+ \\
	&2 M_2^2)+f_{02} f_{12}-f_{02}^2+2 M_0^2 (f_{12}+2 M_2^2))) (-f_{01} (f_{12} f_{23}+2 M_2^2 f_{13})-f_{02} (f_{12} f_{13}+ \\
	&2 M_1^2 f_{23})+f_{03} (f_{12}^2-4 M_1^2 M_2^2)))\Big)
\end{align*}

\section{The generating  function for denominators }\label{sec6}

Now we consider the general generating function \eref{gen-a-1}  for both arbitrary tensor in the numerator and arbitrary  power
distribution  of propagators in
the denominator.
As discussed in section \ref{sec2},  the computation of the generating function \eref{gen-1-1}  can be reduced to the study of the reduction of scalar integral ${{J}}_{n}(t)$\footnote{The notation in \eref{Jt-0} is slightly different from the one in  \eref{gen-a-1} with the region of index: in previous sections it is from $0$ to $n$, while in this section it is from $1$ to $n$.}~\footnote{The reduction of \eref{Jt-0} has also been studied independently in \cite{LZS}.}
\be
J_{n}(\bm{t})\equiv \int {d^D\ell\over i\pi^{D/2}}{1\over \prod_{i=1}^n ((\ell-K_i)^2-M_i^2-t_i)}~.~~\label{Jt-0}
\ee
The IBP relations are given by (see \eref{gen-T-3-3} with $R=0$)
\be {\d\over \d t_i}J_n(\bm{t})=\sum_j (F^{-1})_{ij}(\bm{t}) \left\{ -(D-1-n)J_n(\bm{t})+\sum_{r\neq j} {\d\over \d t_r}J_{n;\WH j}(\bm{t}) \right\}~,~~~~i=1,...,n~\label{Jt-IBP}\ee
where the element of the matrix $F(\bm{t})$ is given by
\be F_{ij}(\bm{t})=(K_i-K_j)^2-M_i^2-t_i-M_j^2-t_j~~~~\label{Ft}\ee
and
\be
J_{n;\WH S}(\bm{t})\equiv \int {d^D\ell\over i\pi^{D/2}}{1\over \prod_{i=1, i\not\in S}^n ((\ell-K_i)^2-M_i^2-t_i)}~.~~\label{Jt-1}
\ee
Using the expansion \eref{gen-a-2-1}, i.e., $ J_{n;\WH S_2}(\bm{t})=\sum_{S_1} J_{n;\WH S_1} {\cal G}_{\WH S_1,\WH S_2}(\bm{t})$, \eref{Jt-IBP} gives following partial differential equations
\be {\d\over \d t_i}{\cal G}_{\WH S_1,\emptyset}(\bm{t})=\sum_j (F^{-1})_{ij}(\bm{t}) \left\{ -(D-1-n){\cal G}_{\WH S_1,\emptyset}(\bm{t})+\sum_{r\neq j} {\d\over \d t_r}{\cal G}_{\WH S_1,\WH j}(\bm{t}) \right\}~.~\label{Jt-IBP-1}\ee
There are two different logics to solve differential equations \eref{Jt-IBP-1}.
The first logic uses iterative construction. Assuming  the general generating functions for $m$-gon with $m<n$ are known, i.e., ${\cal G}_{\WH S_1,\WH j}(\bm{t})
$ is known, we can utilize  \eref{Jt-IBP-1} 
to solve the unknown reduction coefficient ${\cal G}_{\WH S_1,\emptyset}(\bm{t})$. 
We call this logic the "bottom-up" method. The second logic is called the "top-down" method,
which has been demonstrated in \cite{Hu:2023mgc}. In the first step, we solve the case $S_1=\emptyset$ of \eref{Jt-IBP-1}. The result gives the expression ${\cal G}_{\WH i,\WH j}$ with a proper redefinition of the modified Caylay matrix. With this knowledge, we can solve the case $S_1=i$ of \eref{Jt-IBP-1}. By repeating the process we can get the
general result of ${\cal G}_{\WH S_1,\emptyset}$. In fact, in the second subsection, we 
utilize this logic to solve ${\cal G}_{\WH n,\emptyset}$.

Another remark is that compared to the
ordinary differential equation \eref{gen-IBP-3-diff-10} with only one variable, now we have $n$ variables, so solving them
will be difficult in general. However, as we will show,  the one-loop case is special and solving them is relatively easier.

\subsection{The reduction from $n$-gon to $n$-gon}

To demonstrate the key idea of solving \eref{Jt-IBP-1},
let us consider the simplest case $S_1=\emptyset$, i.e., the reduction from $n$-gon to $n$-gon. The equation \eref{Jt-IBP-1} is simplified
to the homogenous equation
\be {\d\over \d t_i}{\cal G}_{\emptyset,\emptyset}(\bm{t})=-\sum_j (F^{-1})_{ij}(\bm{t}) (D-1-n){\cal G}_{\emptyset,\emptyset}(\bm{t})~.~\label{Jt-IBP-nton-1}\ee
We can combine \eref{Jt-IBP-nton-1} to give
\bea \sum_i\left({\d\over \d t_i}{\cal G}_{\emptyset,\emptyset}(\bm{t})\right)dt_i &= & -(D-1-n){\cal G}_{\emptyset,\emptyset}(\bm{t})\sum_{i,j} (F^{-1})_{ij}(\bm{t}) dt_i\nn
& = & -{(D-1-n)\over 2}{\cal G}_{\emptyset,\emptyset}(\bm{t})\sum_{i,j} (F^{-1})_{ij}(\bm{t}) (dt_i+dt_j)\nn
& = & {(D-1-n)\over 2}{\cal G}_{\emptyset,\emptyset}(\bm{t})\sum_{i,j} (F^{-1})_{ij}(\bm{t}) d F_{ij}(\bm{t})~,~\label{Jt-IBP-nton-2}\eea
where we have used the symmetry of $F$-matrix and the expression \eref{Ft}.
To continue, let us recall the following result: for matrix $g_{\mu\nu}$ and determinant $g=|g_{\mu\nu}|$ we have
\be dg= g \left(g^{-1}\right)^{\mu\nu}dg_{\mu\nu}~.~~~\label{matrix-1-1-1}\ee
Thus \eref{Jt-IBP-nton-2} becomes
\be d{\cal G}_{\emptyset,\emptyset}(\bm{t})
 =  {(D-1-n)\over 2}{\cal G}_{\emptyset,\emptyset}(\bm{t}){d|F|\over |F|}~,~\label{Jt-IBP-nton-3}\ee
which can be rewritten as
\be d\ln {\cal G}_{\emptyset,\emptyset}(\bm{t})
 =  {(D-1-n)\over 2}d\ln |F|~.~\label{Jt-IBP-nton-4}\ee
Now we can integrate \eref{Jt-IBP-nton-4} easily to reach
\be
\ln {\cal G}_{\emptyset,\emptyset}(\bm{t})-\ln {\cal G}_{\emptyset,\emptyset}(\bm{t}=0)
 =  {(D-1-n)\over 2}d\ln |F|(\bm{t})-{(D-1-n)\over 2}d\ln |F|(\bm{t}=0)~.~\label{Jt-IBP-nton-5} \ee
Using the boundary condition ${\cal G}_{\emptyset,\emptyset}(\bm{t}=0)=1$ we finally obtain
\be {\cal G}_{\emptyset,\emptyset}(\bm{t})=\left( {|F|(\bm{t})\over |F|(\bm{t}=0)}\right)^{{(D-1-n)\over 2}}~.~\label{Jt-IBP-nton-6}\ee
One can also use \textbf{Lemma 02} given in the Appendix \eqref{sec:prop of F} to check directly that
\eqref{Jt-IBP-nton-6} does satisfy the equation \eref{Jt-IBP-nton-1}. It is very nice to see that
under this construction,  the generating function from $n$-gon to $n$-gon is surprisingly concise.
The result \eref{Jt-IBP-nton-6} tells us also an interesting property of ${\cal H}_{n+1;i}$. From the relation  \eref{gen-a-4}, one
can see that ${\cal H}_{n+1;i}\sim d\log|F|$ after using the result  \eref{Jt-IBP-nton-3}. Since ${\cal H}_{n+1;i}$ appears naturally
in the differential equation of the basis, the dlog-form of ${\cal H}_{n+1;i}$ gives good guidance for the canonical basis. 
\subsubsection{Examples}

Here we give some examples for the result \eref{Jt-IBP-nton-6}. For tadpole
\begin{equation}
	I_1(a)=\int \frac{d^Dl}{i\pi ^{D/2}}\frac{1}{(l^2-M^2)^a}.
\end{equation}
it is easy to find the IBP relation
\begin{equation}
	I_1(a)=\frac{D-2a+2}{2(a-1)M^2}I_1(a-1),
\end{equation}
From it, we can solve the reduction coefficient as
\begin{equation}
	 C_1(a)=\frac{(-1)^{a-1}(1-\frac{D}{2})_{a-1}}{(a-1)!(M^2)^{a-1}}.~~\label{Jt-IBP-nton-tad-c}
\end{equation}
The generating function of this example is
\begin{equation}
	{\cal G}_{\emptyset,\emptyset}(t_1)=\left( 1+\frac{t_1}{M^2}\right)^{{(D-2)\over 2}}~~\label{Jt-IBP-nton-tad}
\end{equation}
It is easy to check the coefficients of its Taylor series expansion in terms of $t_1$ exactly match results in \eref{Jt-IBP-nton-tad-c}. It can be seen that, even though the expression for $C_1(a)$ is not complicated, the form of the generating function is still simpler.

The second example is the massless bubble
\begin{equation}
	I_2(a_1,a_2)=\int\frac{d^Dl}{i \pi^{D/2}}\frac{1}{\left(l^2\right)^{a_1}\left((l-K)^2\right)^{a_2}}.
\end{equation}
The IBP relations give
\begin{equation}
	\begin{aligned}
		 &I_{2}(a_1,a_2)=-\frac{1}{(a_2-1)K^2}[(D-2a_1-a_2+1)I_2(a_1,a_2-1)-(a_2-1)I_2(a_1-1,a_2)],\\
		&I_2(a_1,1)=-\frac{D-a_1-1}{(a_1-1)K^2}I_2(a_1-1,1).
	\end{aligned}~~~\label{nton-bub-1}
\end{equation}
The generating function of this example is
\begin{equation}
	 \mathcal{G}_{\emptyset,\emptyset}(t_1,t_2)=\left(\frac{(K^2)^2+2(t_1+t_2)K^2+(t_1-t_2)^2}{(K^2)^2}\right)^{\frac{D-3}{2}}~~~\label{nton-bub-2}
\end{equation}
From it, one can read out the reduction coefficient as
\begin{equation}
	\begin{aligned}
		 &C_2(a_1,a_2)=\frac{2^{a_1+a_1-2}\left(\frac{D-3}{2}\right)_{\lceil\frac{a_1+a_2-2}{2}\rceil}\left(a_1+1-\frac{D}{2}\right)_{a_2-1}\left(a_2+1-\frac{D}{2}\right)_{\lfloor\frac{a_1-a_2}{2}\rfloor}}{(a_1-1)!(a_2-1)!(K^2)^{a_1+a_2-2}},\ \text{for}\ a_1\geq a_2,\\
		&C_2(a_1,a_2)=C_2(a_2,a_1),\ \text{for}\ a_1<a_2.
	\end{aligned}~~~\label{nton-bub-3}
\end{equation}
where $\lfloor x\rfloor$ represents the floor function of $x$, and $\lceil x \rceil$ represents the ceiling function of $x$.
It can be checked that results \eref{nton-bub-3} do solve \eref{nton-bub-1}.

The third example is  the triangle
\begin{equation}
	I_{3}(a_1,a_2,a_3)=\int {d^D\ell\over i\pi^{D/2}}{\frac{1}{(l^2-M_1^2)^{a_1}((l-K_2)^2-M_2^2)^{a_2}((l-K_3)^2-M_3^3)^{a_3}}}~.
\end{equation}
For this example, we give the numerical check in
the Table \ref{tab:3 to 3}, where the comparison between results calculated using generating functions and those obtained with the Kira program \cite{Maierhofer:2017gsa}. The numerical values we have chosen are
\begin{equation}
	\begin{aligned}
		&D=6.32,\ M^2_1=1.25,\ M^2_2=2.56,\ M^2_3=3.57,\\
		&K_1=\vec{0},\ K^2_2=1.38,\ K_2\cdot K_3=2.05,\ K^2_3=5.16.
	\end{aligned}
	\label{eq:num 01}
\end{equation}

\begin{table}[htbp]\centering\begin{tabular}{|c|c|c|}
		\hline
		$(a_1,a_2,a_3)$ & \text{The numerical result of Kira} & \text{The numerical result of generating function}\\
		\hline
		(3,2,2) & -11.619 & -11.619 \\
		\hline
		(5,2,1) & 19.9606 & 19.9606 \\
		\hline
		(3,3,1) & 9.81226 & 9.81226 \\
		\hline
		(4,3,1) & -32.4261 & -32.4261 \\
		\hline
		(2,2,4) & 11.1596 & 11.1596 \\
		\hline
		(1,6,1) & 2.03967 & 2.03967 \\
		\hline
		(2,1,4) & 1.36335 & 1.36335 \\
		\hline
		(3,5,6) & -193588 & -193588 \\
		\hline
		(6,4,2) & 27906.2 & 27906.2\\
		\hline
	\end{tabular}\caption{Reduction Result of Triangle to Triangle by \texttt{Kira} and generating function}\label{tab:3 to 3}\end{table}

\subsection{ The reduction from $n$-gon to $(n-1)$-gon}

Now we use \eqref{Jt-IBP-1} to find the generating function from $n$-gon to $(n-1)$-gon. For simplicity, let us focus on the case $S_1 = n$. For this case, \eqref{Jt-IBP-1} is not a homogenous differential equation anymore.  Due to $\mathcal{G}_{\widehat{n}, \widehat{j}}(\boldsymbol{t})=0$ for $j\neq n$, there is only one non-zero term in the non-homogeneous part. Expressing the inverse
matrix of $F(\boldsymbol{t})$ in terms of its adjoint matrix $F^*(\boldsymbol{t})$,  \eqref{Jt-IBP-1} becomes
\begin{equation}
	\frac{\partial}{\partial t_i} \mathcal{G}_{\widehat{n}, \emptyset}(\boldsymbol{t})=\left\{-\frac{\sum_j F^{*}_{i j}(\boldsymbol{t})}{|F|(\boldsymbol{t})}(D-1-n) \mathcal{G}_{\widehat{n}, \emptyset}(\boldsymbol{t})\right\}+\frac{F^{*}_{i n}(\boldsymbol{t})}{|F|(\boldsymbol{t})}\sum_{r \neq n} \frac{\partial}{\partial t_r} \mathcal{G}_{\widehat{n}, \widehat{n}}(\boldsymbol{t}),
	\label{eq:diff i n to n-1}
\end{equation}
where $\left(F^*\right)_{i j}(\boldsymbol{t})$ is the algebraic cofactor of the element $F_{ij}(\boldsymbol{t})$ in the matrix $F(\boldsymbol{t})$. Since $\mathcal{G}_{\widehat{n}, \widehat{n}}(\boldsymbol{t})$ gives the reduction of $(n-1)$-gon to $(n-1)$
gon, using the result  \eqref{Jt-IBP-nton-6} we learn
\be \mathcal{G}_{\widehat{n}, \widehat{n}}(\boldsymbol{t})=\left( {F^*_{nn}(\bm{t})\over F^*_{nn}(\bm{t}=0)}\right)^{{(D-n)\over 2}}.~~~\label{Gt-nn}\ee
Now we will solve $N$ partial differential equations
\eref{eq:diff i n to n-1} one by one.

Let us start from the equation $i=n$. Using the method of constant variation, the general solution of the following first-order partial differential equation
\begin{equation}
	\frac{\partial}{\partial x} f(x, \vec{y})=a(x, \vec{y}) f(x, \vec{y})+h(x, \vec{y}),
\end{equation}
is given by
\begin{equation}
	f(x, \vec{y})=e^{\int a(x, \vec{y}) d x}\left(c(\vec{y})+\int e^{-\int a(x, \vec{y}) d x} \cdot h(x, \vec{y}) d x\right) ,
	\label{eq:GS of diff}
\end{equation}
where $\vec{y}$ represents all other variables except $x$. Back to our goal \eref{eq:diff i n to n-1}, first we notice that
the $e^{\pm \int a(x, \vec{y}) d x}$ part is nothing, but $|F|(\boldsymbol{t})^{\pm\frac{D-1-n}{2}}$. This claim  can be derived
by the exact same computations done in the previous subsection.
Secondly, the second term within braces in equation \eqref{eq:GS of diff} is
\bea
	g_{\widehat{n},\emptyset}(\boldsymbol{t})&\equiv	&\int dt_n\left(|F|(\boldsymbol{t})^{-\frac{D-1-n}{2}-1} \cdot F^*_{n n}(\boldsymbol{t}) \cdot \sum_{r \neq n} \frac{\partial}{\partial t_r} \mathcal{G}_{\widehat{n}, \widehat{n}}(\boldsymbol{t})\right)\nn
		& =&F^*_{n n}(\boldsymbol{t}) \cdot \sum_{r \neq n} \frac{\partial}{\partial t_r} \mathcal{G}_{\widehat{n}, \widehat{n}}(\boldsymbol{t})\mathcal{T}(\boldsymbol{t}), \nn
		\mathcal{T}(\boldsymbol{t})&\equiv &\int dt_n\left(|F|(\boldsymbol{t})^{-\frac{D-1-n}{2}-1}\right)
		\label{eq:T of n-1}
\eea
where we have used the fact  that  $F^*_{n n}(\boldsymbol{t}) \cdot \sum_{r \neq n} \frac{\partial}{\partial t_r} \mathcal{G}_{\widehat{n}, \widehat{n}}(\boldsymbol{t})$  is independent of $t_n$. The part that needs to be integrated is only $|F|(\boldsymbol{t})^{-\frac{D-1-n}{2}-1}$. Using results given in \textbf{Lemma 01} and \textbf{Lemma 02} in the Appendix \ref{sec:prop of F}, we have
\begin{equation}
	\begin{aligned}
		|F|(\boldsymbol{t})=a_0+a_1 t_n +a_2t_n^2
	\end{aligned}
\end{equation}
with coefficients $a_0,a_1,a_2$
\be
		a_0=|F|(\boldsymbol{t})\big|_{t_n=0},~~~~
		a_1=-2\sum_{i=1}^n F^*_{ni}(\boldsymbol{t})\big|_{t_n=0},~~~~
		a_2=-\sum_{i\neq n,j\neq n} F^*_{in,jn}(\boldsymbol{t}).
\ee
While $a_0$ and $a_1$ depend on $t_1,t_2,\cdots, t_{n-1}$, $a_2$ is a pure constant. Then according to the integral formula,
\begin{equation}
	\int dx\  \left((x-x_+)(x-x_-)\right)^c ={1\over 1+c}\left(x-x_+\right)^{c+1}\left(x-x_-\right)^c\cdot\  _2F_1\left(1,-c;2+c;\frac{x-x_+}{x-x_-}\right),
\end{equation}
integral in \eqref{eq:T of n-1} is
\be	\mathcal{T}(\boldsymbol{t})={-2\over D-n-1}	 a_2^{-\frac{D-n+1}{2}}(t_n-x_+)^{-\frac{D-n-1}{2}}(t_n-x_-)^{-\frac{D-n+1}{2}}\cdot \ _2F_1\left(1,\frac{D-n+1}{2};\frac{n+3-D}{2};\frac{t_n-x_+}{t_n-x_-}\right)
	\label{eq:sp solution n-1}
\ee
with $x_\pm=\frac{-a_1\pm \sqrt{a^2_1-4a_0a_2}}{2a_2}$. Putting it back we have
\begin{equation}
	\begin{aligned}
		g_{\widehat{n},\emptyset}(\textbf{t})=&{-2\over D-n-1}\left(\left(F^*_{n n}(\boldsymbol{t})\right) \cdot \sum_{r \neq n} \frac{\partial}{\partial t_r} \mathcal{G}_{\widehat{n}, \widehat{n}}(\boldsymbol{t})\right)a_2^{-\frac{D-n+1}{2}}(t_n-x_+)^{-\frac{D-n-1}{2}}(t_n-x_-)^{-\frac{D-n+1}{2}}\\
		\times & _2F_1\left(1,\frac{D-n+1}{2};\frac{n+3-D}{2};\frac{t_n-x_+}{t_n-x_-}\right).
	\end{aligned}
	\label{eq:solu g}
\end{equation}

 Up to now the generating function $\mathcal{G}_{\widehat{n},\emptyset}(\boldsymbol{t})$ from $n$-gon to $(n-1)$-gon is
\begin{equation}
	\begin{aligned}
		\mathcal{G}_{\widehat{n},\emptyset}(\boldsymbol{t})=&{-2\over (D-n-1)a_2}\cdot \frac{1}{t_n-x_-}\ _2F_1\left(1,\frac{D-n+1}{2};\frac{n+3-D}{2};\frac{t_n-x_+}{t_n-x_-}\right)\left(\left(F^*_{n n}(\boldsymbol{t})\right) \cdot \sum_{r \neq n} \frac{\partial}{\partial t_r} \mathcal{G}_{\widehat{n}, \widehat{n}}(\boldsymbol{t})\right)\\
		+&c_1(t_1,\cdots,t_{n-1})\cdot \left(|F|(\boldsymbol{t})\right)^{\frac{D-1-n}{2}}.
	\end{aligned}
	\label{eq:c n to n-1}
\end{equation}
To solve the undetermined function $c(t_1,\cdots,t_{n-1})$, we can substitute the above expression into the differential equation  \eqref{eq:diff i n to n-1} with $i=n-1$. Doing similar computations, we will find expressions with the undetermined function $c_2(t_1,\cdots,t_{n-2})$ in principle.
Iterating the procedure until $i=1$ and using the boundary condition we should be able to
fix $\mathcal{G}_{\widehat{n},\emptyset}(\boldsymbol{t})$ completely.
 However, in the Appendix \ref{sec:proof n to n-1} we will prove that \textbf{ the $c_1(t_1,\cdots,t_{n-1})$ in expression \eqref{eq:c n to n-1} is actually a constant independent of $t_1,\cdots, t_{n-1}$}. Therefore, using the boundary conditions, the expression for the very concise generating function $\mathcal{G}_{\widehat{n},\emptyset}(\boldsymbol{t})$ from $n$-gon to $(n-1)$-gon is
\begin{equation}
	\begin{aligned}
		 \mathcal{G}_{\widehat{n},\emptyset}(\boldsymbol{t})=&\frac{-2}{(D-n-1)a_2}\frac{1}{t_n-x_-}\ _2F_1\left(1,\frac{D-n+1}{2};\frac{n+3-D}{2};\frac{t_n-x_+}{t_n-x_-}\right)\left(F^*_{n n}(\boldsymbol{t}) \cdot \sum_{r \neq n} \frac{\partial}{\partial t_r} \mathcal{G}_{\widehat{n}, \widehat{n}}(\boldsymbol{t})\right)\\
		-&g_{\widehat{n},\emptyset}(\boldsymbol{t}=0)\cdot \left(|F|(\boldsymbol{t})\right)^{\frac{D-1-n}{2}}.
		\label{eq: GF n to n-1}
	\end{aligned}
\end{equation}
where $g_{\widehat{n},\emptyset}(\boldsymbol{t}=0)$ is the numerical value obtained by setting all $t_i$ to zero in equation \eqref{eq:solu g}. Substituting the expression of $\mathcal{G}_{\widehat{n},\widehat{n}}(\boldsymbol{t})$, the above formula can be further simplified as:
\begin{equation}
	\begin{aligned}
		 \mathcal{G}_{\widehat{n},\emptyset}(\boldsymbol{t})=&-\frac{2(D-n)}{D-n-1}\Bigg(\frac{1}{t_n-x_-}\ _2F_1\left(1,\frac{D-n+1}{2};\frac{n+3-D}{2};\frac{t_n-x_+}{t_n-x_-}\right)\left( {F^*_{nn}(\bm{t})\over F^*_{nn}(\bm{t}=0)}\right)^{{(D-n)\over 2}}\\
		-&\left\{\frac{1}{t_n-x_-}\ _2F_1\left(1,\frac{D-n+1}{2};\frac{n+3-D}{2};\frac{t_n-x_+}{t_n-x_-}\right)\right\}\Bigg|_{\boldsymbol{t}\to 0}\left( {|F|(\bm{t})\over |F|(\bm{t}=0)}\right)^{{(D-1-n)\over 2}}\Bigg).
		\label{eq: GF n to n-1}
	\end{aligned}
\end{equation}

\subsubsection{Numerical check}
Here we  present the numerical results of the triangle reduction to verify our calculations. We choose the same numerical values as \eqref{eq:num 01}. Table \ref{tab:3 to 2-1} and Table \ref{tab:3 to 2-2} list the comparison between the results calculated using generating functions and those obtained with the Kira program.

\begin{table}[htbp]\centering\begin{tabular}{|c|c|c|}
		\hline
		$(a_1,a_2,a_3)$ & \text{The numerical result of Kira} & \text{The numerical result of generating function}\\
		\hline
		(3,2,2) & 10.2631 & 10.2631 \\
		\hline
		(5,2,1) & -17.6323 & -17.6323 \\
		\hline
		(3,3,1) & -8.65731 & -8.65731 \\
		\hline
		(4,3,1) & 28.6463 & 28.6463 \\
		\hline
		(2,2,4) & -9.85984 & -9.85984 \\
		\hline
		(1,6,1) & -1.80302 & -1.80302 \\
		\hline
		(2,1,4) & -1.20463 & -1.20463 \\
		\hline
	\end{tabular}\caption{Reduction Result of Triangle to $D_1,D_2$ Bubble by \texttt{Kira} and generating function}\label{tab:3 to 2-1}\end{table}

\begin{table}[htbp]\centering\begin{tabular}{|c|c|c|}
		\hline
		$(a_1,a_2,a_3)$ & \text{The numerical result of Kira} & \text{The numerical result of generating function}\\
		\hline
		(3,2,2) & 7.67615 & 7.67615 \\
		\hline
		(5,2,1) & -13.1893 & -13.1893 \\
		\hline
		(3,3,1) & -6.48029 & -6.48029 \\
		\hline
		(4,3,1) & 21.4263 & 21.4263 \\
		\hline
		(2,2,4) & -7.37398 & -7.37398 \\
		\hline
		(1,6,1) & -1.34816 & -1.34816 \\
		\hline
		(2,1,4) & -0.900785 & -0.900785 \\
		\hline
	\end{tabular}\caption{Reduction Result of Triangle to $D_2,D_3$ Bubble by \texttt{Kira} and generating function}\label{tab:3 to 2-2}\end{table}

\subsection{The reduction from $n$-gon to general $m$-gon}

Now we solve the differential equations of  \eref{Jt-IBP-1} for general cases.
Mimic the derivation of \eref{Jt-IBP-nton-2} we get
\bea d{\cal G}_{\WH S_1,\emptyset}(\bm{t})
& = & {(D-1-n)\over 2}{\cal G}_{\WH S_1,\emptyset}(\bm{t}){d\ln |F|} +\sum_i {\cal B}_{i;\WH S_1}(\bm{t})dt_i\nn
{\cal B}_{i;\WH S_1}(\bm{t}) & = &\sum_j (F^{-1})_{ij}(\bm{t}) \sum_{r\neq j} {\d\over \d t_r}{\cal G}_{\WH S_1,\WH j}(\bm{t})  ~.~\label{Jt-IBP-g-1}\eea
By the constant variation method, we write
\be {\cal G}_{\WH S_1,\emptyset}(\bm{t})= \left( {|F|(\bm{t})\over |F|(\bm{t}=0)}\right)^{{(D-1-n)\over 2}}\W g_{\WH S_1,\emptyset}(\bm{t}) ~.~\label{Jt-IBP-g-2}\ee
Putting it back to \eref{Jt-IBP-g-1} and doing some manipulations we get
\be d{\W g}_{\WH S_1,\emptyset}(\bm{t})=\left( {|F|(\bm{t})\over |F|(\bm{t}=0)}\right)^{-{(D-1-n)\over 2}}\sum_i {\cal B}_{i;\WH S_1}(\bm{t})dt_i ~.~\label{Jt-IBP-g-5}\ee
To solve ${\W g}_{\WH S_1,\emptyset}(\bm{t})$, we just need to do multi-variable integrations. Noticing that  the existence of solution of \eref{Jt-IBP-g-5} means the integrability condition
\bea d\left\{\left( {|F|(\bm{t})\over |F|(\bm{t}=0)}\right)^{-{(D-1-n)\over 2}}\sum_i {\cal B}_{i;\WH S_1}(\bm{t})dt_i\right\}=0~,~\label{Jt-IBP-g-6}\eea
thus we have the freedom to choose integration paths.
The procedure done in previous subsection is, in fact, equivalent to following choice of integration path, i.e., beginning at point $(0,0,\cdots,0)$, proceeding to point $(0,0,\cdots,t_n)$, then to point $(0,\cdots,t_{n-1},t_n)$, and so forth, until finally reaching the integration endpoint $(t_1,t_2,\cdots,t_n)$. Another choice is the path that connecting point $(0,0,\cdots,0)$ to $(t_1,t_2,\cdots,t_n)$ by a straight line. This line is parameterized as $(\lambda t_1,\lambda t_2,\cdots ,\lambda t_n)$, thus
the solution of \eref{Jt-IBP-g-5} is
\bea {\W g}_{\WH S_1,\emptyset}(\bm{t})-{\W g}_{\WH S_1,\emptyset}(\bm{t}=0)=\int_0^1 d\lambda \left( {|F|(\lambda\bm{t})\over |F|(\bm{t}=0)}\right)^{-{(D-1-n)\over 2}}\sum_i {\cal B}_{i;\WH S_1}(\lambda\bm{t}) t_i ~~\label{Jt-IBP-g-7}\eea
where the boundary condition is that
\bea {\W g}_{\WH S_1,\emptyset}(\bm{t}=0)=\left\{ \begin{array}{ll} 1,~~ & S_1=\emptyset \\
	0,~~ & S_1\neq \emptyset
\end{array}\right.~~\label{Jt-IBP-g-8}\eea
In principle, knowing the general solution \eref{Jt-IBP-nton-6} from $n$-gon to $n$-gon, one can use \eref{Jt-IBP-g-7}
to get the general solution from $n$-gon to $(n-1)$-gon. Iterating the procedure, one can find the general solution from
$n$-gon to arbitrary $m$-gon, although carrying out the integration explicitly will need some work.

Results in this section, including \eref{Jt-IBP-nton-6},  \eref{eq: GF n to n-1}, \eref{Jt-IBP-g-2} and   \eref{Jt-IBP-g-7},
are the fourth main result in the paper. With the explicit compact generating functions, no matter whether analytically or numerically
the computation of reduction coefficients becomes much simpler. 
\section{Conclusion}

In this paper, we have studied the general generating function for the reduction of one-loop integrals with arbitrary tensor structure in the numerator
and arbitrary power distribution of propagators in the denominator. Our goal is to find the compact analytical expressions 
for these generating functions. To achieve this goal, we have separated the task into two parts:
the reduction with only an arbitrary numerator and the reduction of scalar integrals with an arbitrary denominator.
Using the IBP method, we have derived corresponding differential equations for these two generating functions.
By solving these equations, we obtained the following four main results of the paper: (1) An analytic  recursive computation for generating functions of the tensor reduction given in \eref{gen-IBP-3-5-10b}, which is easy to implement in Mathematica; (2) A compact analytical closed forms for generating functions of
the tensor reduction
given in \eref{gen-IBP-3-diff-11} and \eref{gen-IBP-3-diff-12}; (3) The recursive algorithm for the matrix ${\cal H}_{n+1;i}$ 
(see \eref{gen-T-H-1}, \eref{H-algo-1} and \eref{H-algo-2})
describing the reduction of basis under the action of ${\partial \over \partial M^2}$, which can easily be  implemented  in  Mathematica; (4) A compact analytical closed forms for generating functions of
the denominator reduction given in \eref{Jt-IBP-nton-6},  \eref{eq: GF n to n-1}, \eref{Jt-IBP-g-2} and   \eref{Jt-IBP-g-7}.

With the obtained analytic expressions of generating functions, we can improve the efficiency of computations of loop integrals. For example, from the expression, we can get information on the singularities of reduction coefficients when doing the phase-space integration. Using this information we can adjust the algorithm to make the numerical evaluation more stable. Another benefit is that 
if we put the numerical external data into the generating function, the numerical evaluation of reduction coefficients will be faster. 
A third possible application is as follows. Recently, the functional reconstruction with finite fields \cite{vonManteuffel:2014ixa, Peraro:2016wsq} has been applied to the reduction 
of higher loop integrals and has been proven to be a powerful method. The analytic structure of reduction coefficients, for example,
the factor of denominators, will make the reconstruction faster.

The knowledge of reduction coefficients is also useful for the
evaluation of master integrals. For example, since the parameters ${\bm t}$ have played the role of the mass if we set
$M_i=0$ in \eref{Jt-0}, the generating function ${\cal G}$ studied in the section \ref{sec6} will give the expansion
of massive scalar integrals by the massless integrals.
Another example is the following. Currently, the standard way to compute master integrals is to use the differential equation \cite{Kotikov:1990kg, Kotikov:1991pm, Bern:1993kr, Gehrmann:1999as}, i.e.,
finding
\bea {\d \over \d s}{\cal I}=A_s {\cal I}~~~\label{diff-I}\eea
When acting  ${\d \over \d s}$ on the master integrals, we get integrals with nontrivial numerators and higher power of propagators at the left-hand side of \eref{diff-I}. After reducing them to the basis, we get the right-hand side of \eref{diff-I}.
Since the generating function contains all information of reduction,  one could read out the matrix $A_s$ directly
from it. Furthermore, one could try to find canonical basis \cite{Henn:2013pwa} as well as the symbol \cite{Brown:2009qja, Goncharov:2009lql, Goncharov:2010jf} directly using the generating function.
Working out these details will be interesting.

In this paper, we have used only fundamental IBP relations to establish differential equations. As demonstrated
in \cite{Gluza:2010ws, Larsen:2015ped, Ita:2015tya} (i.e., the module intersection method), we could use non-fundamental IBP relations to avoid the appearance of propagators with higher powers. In fact, Kosower has used exactly the same idea to discuss the
generating function for some particular tensor structures in \cite{Kosower:2018obg}. While the module intersection method will
simplify resulted IBP relations, finding the syzygy does take some considerable computations. It will be interesting to
investigate when applying the same idea to generating functions, which kinds of simplification we will get. Furthermore, there are other reduction methods, such as the unitarity cut method,
intersection theory method or recently proposed linear relation method in \cite{Liu:2022mfb}. Thus it is natural to ask
which insight we will obtain if using them to study the generating function\footnote{In \cite{Guan:2023avw} the generating function has been studied using the linear relation method.}.

Finally, as we have mentioned in the introduction, one of our main motivations is to generalize the generating function to higher loop integrals. 	Although the experience of the one-loop in this paper will be very helpful, we expect more complicated technical problems will appear for higher loops. Solving them will be an exciting  challenge.

\section*{Acknowledgements}

We would like to thank Yang Zhang, Tingfei Li, Xiang Li and Yongqun Xu for the useful discussion. Chang Hu and Jiyuan Shen thank  Hangzhou Institute for Advanced Study for financial support.
Yaobo Zhang thanks Ningxia University for providing the startup fund for young faculty research. This work is supported by Chinese NSF funding under Grant No.11935013, No.11947301, No.12047502 (Peng Huanwu Center) and NSAF grant in National Natural Science Foundation of China (NSFC) with grant No. U2230402.

\appendix

\section{Solving the differential equation}

In this appendix, we give some details on solving the differential equation for self-contained use.

\subsection{The generalized hypergeometric functions}

First, we collect some mathematical results which are used in the paper.
The {\bf generalized hypergeometric function} (see (C.2) of \cite{Weinzierl:2022eaz} or the section 2.9 of
\cite{MOS}) is defined as
\be ~_A F_B (a_1,...,a_A; b_1,...,b_B;x)=\sum_{n=0}^\infty { (a_1)_n ... (a_A)_n \over (b_1)_n ...(b_B)_n} {x^n\over n!}~\co~~~\label{Spe-1-1}\ee
where  the  {\bf Pochhammer symbol} is given in \eref{Poch-1}.
Let $\theta\equiv x{d\over d x}$,
then $w=~_A F_B$ satisfies the differential equation
\be \left( \theta(\theta+b_1-1)...(\theta+b_B-1)-x (\theta+a_1)...(\theta+a_A)\right) w=0~.~~\label{Spe-1-3}\ee
Let us consider the special case. When $A=B=1$, the solution is called the {\bf Kummer's function}. It is easy to work out that the differential equation \eref{Spe-1-3} is
	\be x\left(x{d^2\over dx^2}+(b_1-x){d\over dx}-a_1\right) w=0~~~\label{Spe-2-1}\ee
	%
	and the solution is $~_1 F_1 (a_1; b_1;x)$.
	If we rescale $x=\a z$, \eref{Spe-2-1} will becomes
	\be \left( z {d^2\over dz^2}+{(b_1-\a z)}{d\over dz}-a_1 \a\right)w=0~~~\label{Spe-2-1a}\ee
	and the solution will be  $~_1 F_1 (a_1; b_1;\a z)$.

\subsection{The inhomogeneous differential equation}

Now we consider the inhomogeneous differential equation of the following type
\be \left( x{d^2\over dx^2}+(b_1-x){d\over dx}-a_1\right)w=g(x)~.~~\label{Spe-2-4}\ee
Using the result \eref{Spe-2-1}, we write
\be w= ~_1F_1(a_1;b_1;x)\W H(x)~,~~\label{Spe-2-5}\ee
thus we have the differential equation for $\W H(x)$
\be x ~_1F_1 \W H''+\W H'(2x ~_1F'_1+(b_1-x)~_1F_1)=g(x)~.~~\label{Spe-2-6}\ee
We can further simplify by defining
\bea  \W H' &= & h(x) e^{-\int_0^x dt {(2t ~_1F'_1(a_1;b_1;t)+(b_1-t)~_1F_1(a_1;b_1;t))\over t ~_1F_1(a_1;b_1;t) }}\nn
& = & h(x) e^{-\left( 2  \ln (~_1F_1(a_1;b_1;x))+b_1\ln x-x\right)}= {h(x) x^{-b_1} e^x\over (~_1F_1(a_1;b_1;x))^2} ~.~~\label{Spe-2-7}\eea
Putting \eref{Spe-2-7} to \eref{Spe-2-6} we find
\be h'=g(x) x^{b_1-1} ~_1F_1(a_1;b_1;x)e^{-x}~,~~\label{Spe-2-8}\ee
thus we can solve
\be h(x)=C_2+ \int_0^x d y g(y) y^{b_1-1} ~_1F_1(a_1;b_1;y)e^{-y}~~~\label{Spe-2-9}\ee
and
\be \W H(x)  =  C_1+\int_0^x dt { t^{-b_1} e^t\over (~_1F_1(a_1;b_1;t))^2}
\left(C_2+ \int_0^t d y g(y) y^{b_1-1} ~_1F_1(a_1;b_1;y)e^{-y}\right)~.~~\label{Spe-2-13}\ee
Finally, we have the solution\footnote{In fact, to require the finiteness at $x=0$, one must take $C_2=0$.
	Furthermore, \eref{Spe-2-14} can have a nice formal expression as discussed in \eref{ser-2-4}.}
\bea w=~_1F_1(a_1;b_1;x)\left\{ C_1+\int_0^x dt { t^{-b_1} e^t\over (~_1F_1(a_1;b_1;t))^2}
\left(C_2+ \int_0^t d y g(y) y^{b_1-1} ~_1F_1(a_1;b_1;y)e^{-y}\right)\right\}.~~~~~\label{Spe-2-14}\eea
In \eref{Spe-2-14} there are two unknown constants $C_1, C_2$ which is the typical character for
second order differential equation. We should use the boundary condition to determine them.

\subsection{Solving equation  }

Now we use the result in previous subsections to solve equation \eref{gen-IBP-3-diff-11}, which, after proper
scaling,  can be
simplified as
\be t {d^2\over d t^2}  \vec{W}+ (B_0+B_1 t){d\over dt}\vec{W}+(C_0+C_1 t)\vec{W}+ \vec{\cal B}(t)=0~~~~\label{Diff-1-1}\ee
To solve it, we need to do the following two steps.  Since \eref{Diff-1-1} is not the standard form \eref{Spe-2-1a} yet, we do the
rescaling
\be \vec{W}= f(t) \vec{U}~~~\label{Diff-2-1}\ee
with
\be f''+B_1f' +C_1 f=0,~~~\Longrightarrow f= e^{{-B_1+\sqrt{B_1^2-4 C_1}\over 2}t}~~~\label{Diff-2-3}
\ee
to reach the wanted form of differential equation
\bea 0 & = & t {d^2\over d t^2}  \vec{U}+ \left( B_0+t\sqrt{B_1^2-4 C_1}\right){d\over dt}\vec{U}\nn
& & +\left( {-B_1+\sqrt{B_1^2-4 C_1}\over 2}B_0+ C_0 \right)\vec{U}+ e^{{B_1-\sqrt{B_1^2-4 C_1}\over 2}t}\vec{\cal B}(t)~~~\label{Diff-2-4}\eea
Let us define
\bea x= -t\sqrt{B_1^2-4 C_1}~~~\label{Diff-2-5}\eea
then the above equation becomes
\bea 0 & = & x {d^2\over d x^2}  \vec{U}+ \left( B_0-x\right){d\over dx}\vec{U}
 +{-1\over \sqrt{B_1^2-4 C_1}}\left( {-B_1+\sqrt{B_1^2-4 C_1}\over 2}B_0+ C_0 \right)\vec{U}\nn & & + {-1\over \sqrt{B_1^2-4 C_1}}e^{{B_1-\sqrt{B_1^2-4 C_1}\over 2}{-x\over \sqrt{B_1^2-4 C_1}}}\vec{\cal B}({-x\over \sqrt{B_1^2-4 C_1}})~~~\label{Diff-2-6}\eea
Comparing it with \eref{Spe-2-4}, we see that
\bea a_1 & = & {1\over \sqrt{B_1^2-4 C_1}}\left( {-B_1+\sqrt{B_1^2-4 C_1}\over 2}B_0+ C_0 \right),~~~~~~
b_1  =  B_0, \nn
g(x) & = & {1\over \sqrt{B_1^2-4 C_1}}e^{{B_1-\sqrt{B_1^2-4 C_1}\over 2}{-x\over \sqrt{B_1^2-4 C_1}}}\vec{\cal B}({-x\over \sqrt{B_1^2-4 C_1}})~~~\label{Diff-2-7}\eea
Thus using \eref{Spe-2-14}, we have
\bea \vec{U}(x) & = & (~_1F_1(a_1;b_1;x))\left\{ c_1+\int_0^x ds { s^{-b_1} e^s\over (~_1F_1(a_1;b_1;s))^2}
\left(c_2+ \int_0^s d y g(y) y^{b_1-1} ~_1F_1(a_1;b_1;y)e^{-y}\right)\right\}~~~\label{Diff-2-8}\eea
and finally
\bea \vec{W}(t) & = & e^{-{B_1+\sqrt{B_1^2-4 C_1}\over 2}t} \vec{U}(x)~~~\label{Diff-2-9}\eea
%

\subsection{Series expansion}

Although we have obtained the closed analytic expression, it is still useful to give the series expansion to demonstrate
the vanishing of
square root for each order of $R$.

Putting the series expansion
\be W=\sum_{n=0}^\infty \vec w_n t^n,~~~~\vec{\cal B}(t)=\sum_{n=0}^\infty \vec b_n t^n~~~\label{ser-1-1}\ee
to  \eref{Diff-1-1}, we get the recursion relation
\be 0  = \vec w_{n+1} (n+1) (n+B_0)+\vec  w_n (B_1 n+C_0)+\vec  w_{n-1} C_1+\vec  b_n~~~\label{ser-2-1}\ee
where $w_{n<0}=0$ has been assumed. When $n=0$, it is reduced to
\be 0  = \vec w_{1} B_0+\vec  w_0 C_0+\vec  b_0~~~\label{ser-2-2}\ee
Using the boundary condition $\vec{w}_0=\vec{\a}_0\equiv \{1,0,0,...,0\}^T$ and $\vec  b_0=0$, we can solve
\be \vec w_{1}={-C_0\over  B_0}\vec  \a_0~~~\label{ser-2-3}\ee
For $n\geq 1$ we have
\be \vec w_{n+1}={-(B_1 n+C_0)\over  (n+1) (n+B_0)}\vec  w_n
+{-C_1\over  (n+1) (n+B_0)}\vec  w_{n-1} +{-1\over  (n+1) (n+B_0)}\vec  b_n~~~\label{ser-2-4}\ee
Thus all $\vec{w}_n$ have been determined.

\section{Some properties about the matrix F(t)}
\label{sec:prop of F}
In this section, we will present some properties of the matrix $F(\boldsymbol{t})$ defined in \eref{Ft}.
Utilizing these properties, we can derive several very useful results for the reduction of $n$-gon to the $(n-1)$-gon. We will not provide proofs for these properties, since readers can easily verify their correctness with basic linear algebra techniques.

\begin{itemize}
	\item \textbf{Lemma 01}: With the explicit form \eref{Ft},
	the determinant of $F(\boldsymbol{t})$ is following form
	\begin{equation}
		|F|(\boldsymbol{t})=\sum_{i\leq j}c_{ij} t_i t_j +\sum_{i=1}^n c_i t_i +c_0.~~~\label{Hu-F-1}
	\end{equation}
	where $c_{ij},c_i,c_0$ are independent of $t_1,\cdots,t_n$.
	
	\item \textbf{Lemma 02}:
	For $i,j=1,2,\cdots ,n$ we have,
	\begin{equation}
		\begin{aligned}
			&\frac{\partial}{\partial t_i}|F|\left(\boldsymbol{t}\right)=-2\left(F_{i 1}^*(\boldsymbol{t})+F_{i 2}(\boldsymbol{t})^*+\cdots+F_{i n}^*(\boldsymbol{t})\right),\\
			&\frac{\partial^2|F|\left(\boldsymbol{t}\right)}{\partial t^2_i}=-\sum_{k_1,k_2\neq i} F^*_{ik_1,ik_1}(\boldsymbol{t}),\\
			&\frac{\partial^2|F|\left(\boldsymbol{t}\right)}{\partial t_i\partial t_j}=+\sum_{k_1\neq i,k_2\neq j} F^*_{ik_1,jk_1}(\boldsymbol{t})\ \  \text{for}\ i\neq j.~~~\label{Hu-F-2}
		\end{aligned}
	\end{equation}
where $F_{i 1}^*$ is the cofactor and $F^*_{ik_1,ik_1}$ the second order cofactor.
More explicitly, if we define $Q^{\#}_{i_1i_2,j_1j_2}$ as the $(n-2)\times(n-2)$ matrix obtained by removing the $i_1$-th row and $i_2$-th row, as well as the $j_1$-th column and $j_2$-th column from the matrix $Q$, then the second order cofactor $Q^*_{i_1i_2,j_1j_2}=(-1)^{i_1+i_2+j_1+j_2}|Q^{\#}_{i_1i_2,j_1j_2}|$. Similarly, we can define the third-order algebraic cofactor $Q_{i_1i_2i_3,j_1j_2j_3}$ as well as the higher-order cofactor.
	
	\item \textbf{Lemma 03}:
	For the algebraic cofactors of matrix $F(\boldsymbol{t})$ and $i\neq j \neq k$, we have
	\begin{equation}
		\begin{aligned}
			&\frac{\partial F_{i j}^*(\boldsymbol{t})}{\partial t_i}+\frac{1}{2} \frac{\partial F_{i i}^*(\boldsymbol{t})}{\partial t_j}=0,\\
			&\frac{\partial F^*_{ij}(\boldsymbol{t})}{\partial t_k}=\sum_{l\neq i} F^*_{il,jk}(\boldsymbol{t}) +\sum_{l\neq j} F^*_{ik,jl}(\boldsymbol{t})\\
			&\frac{\partial}{\partial t_k}\sum_{i,j=1}^n F^*_{ij}(\boldsymbol{t})=0,\ \text{for}\  k=1,2,\cdots,n.
		\end{aligned}
	\end{equation}
	
	\item  \textbf{Lemma 04}:
	For any matrix $A$, and $i_1<i_2$, $j_1<j_2$, The \textbf{Desnanot-Jacobi identity} is given by
	\begin{equation}
		\begin{aligned}
			A_{i_1 j_1}^* A_{i_2 j_2}^*-A_{i_1 j_2}^* A_{i_2 j_1}^*=A_{i_1 i_2, j_1 j_2}^*|A|.
		\end{aligned}
	\end{equation}
\end{itemize}

\section{Proof of the correctness of the generating function from $n$-gon to $(n-1)$-gon}
\label{sec:proof n to n-1}

In this section, we will proof that  the $c_1(t_1,\cdots,t_{n-1})$ in expression \eqref{eq:c n to n-1} is actually a constant independent of $t_1,t_1,\cdots, t_{n-1}$. In fact, this claim is equivalent to the claim that  $g_{\widehat{n},\emptyset}(\boldsymbol{t})$ given by  \eqref{eq:T of n-1} and \eqref{eq:sp solution n-1} satisfies
\begin{equation}
	\frac{\partial}{\partial t_i} g_{\widehat{n}, \emptyset}(\boldsymbol{t})=-|F|(\boldsymbol{t})^{-\frac{D-1-n}{2}-1}\cdot F^{*}_{i n}(\boldsymbol{t})\sum_{r \neq n} \frac{\partial}{\partial t_r} \mathcal{G}_{\widehat{n}, \widehat{n}}(\boldsymbol{t}),
	\label{eq:proof 01}
\end{equation}
for $i=1,2,\cdots,n$. Without loss of generality, we provide the proof for the case $i=1$. We rewrite the determinant of matrix $F(\boldsymbol{t})$ in the following form
\begin{equation}
	\begin{aligned}
		|F|(\boldsymbol{t})=A_1 t_1^2 + A_2 t_n^2 + A_3t_1t_n + A_4 t_1 + A_5 t_n + A_6.
	\end{aligned}
\end{equation}
Then,  utilizing the Euler transformation formula of hypergeometric functions
\begin{equation}
	_2 F_1(a, b ; c ; z)=(1-z)^{-b}\ _2 F_1\left(c-a, b ; c ; \frac{z}{z-1}\right),
\end{equation}
we can rewrite \eqref{eq:sp solution n-1} as
\begin{equation}
	\begin{aligned}
		\mathcal{T}(\boldsymbol{t})=& \frac{2}{1-D+n}\frac{-\sqrt{(A_3 t_1+A_5)^2-4 A_2(A_1 t_1^2+A_4 t_1+A_6)}+2 A_2 t_n+A_3 t_1+A_5}{2 A_2} \\
		\times & \left(\frac{\sqrt{(A_3 t_1+A_5)^2-4 A_2(A_1 t_1^2+A_4 t_1+A_6)}+2 A_2 t_n+A_3 t_1+A_5}{2 \sqrt{(A_3 t_1+A_5)^2-4 A_2(A_1 t_1^2+A_4 t_1+A_6)}}\right)^{\frac{1}{2}(D-n-1)+1} \\
		\times & \left(A_1 t_1^2 + A_2 t_n^2 + A_3t_1t_n + A_4 t_1 + A_5 t_n + A_6\right)^{\frac{1}{2}(-D+n+1)-1} \\
		\times & { }_2 F_1\Bigg(\frac{1}{2}(D-n-1)+1, \frac{1}{2}(-D+n+1) ; \frac{1}{2}(-D+n+1)+1 ;\\
		&\frac{\sqrt{(A_3 t_1+A_5)^2-4 A_2(A_1 t_1^2+A_4 t_1+A_6)}-2 A_2 t_n-A_3 t_1-A_5}{2 \sqrt{(A_3 t_1+A_5)^2-4 A_2(A_1 t_1^2+A_4 t_1+A_6)}}\Bigg).
	\end{aligned}
	\label{eq:proof 02}
\end{equation}
The advantage of this transformation is that we can utilize the following derivative formula for hypergeometric functions
\begin{equation}
	\frac{d}{dx}\ {}_2F_1\left(-c,c+1;c+2;x\right)=\frac{(c+1)\left((1-x)^c-{ }_2 F_1(-c, c+1 ; c+2 ; x)\right)}{x},
\end{equation}
thus it ensures that the parameters of the hypergeometric function remain unchanged during the differentiation process of $g_{\widehat{n},\emptyset}(\boldsymbol{t})$ with respect to $t_1$.
\begin{equation}
	\begin{aligned}
		\frac{\partial\mathcal{T}(\boldsymbol{t})}{\partial t_1}=&\left(-\frac{2(n-D)}{1-D+n}\right) \frac{4A_1A_2t_1+2A_2A_4-A^2_3t_1-A_3A_5}{\left((A_3 t_1+A_5)^2-4 A_2(A_1 t_1^2+A_4 t_1+A_6)\right)^{3 / 2}} \\
		\times & \left(\frac{\sqrt{(A_3 t_1+A_5)^2-4 A_2(A_1 t_1^2+A_4 t_1+A_6)}+2 A_2 t_n+A_3 t_1+A_5}{2 \sqrt{(A_3 t_1+A_5)^2-4 A_2(A_1 t_1^2+A_4 t_1+A_6)}}\right)^{\frac{1}{2}(D-n-1)} \\
		\times & \left(A_1 t_1^2 + A_2 t_n^2 + A_3t_1t_n + A_4 t_1 + A_5 t_n + A_6\right)^{\frac{1}{2}(-D+n+1)}\\
		\times &{ }_2 F_1\Bigg(\frac{1}{2}(D-n-1)+1, \frac{1}{2}(-D+n+1) ; \frac{1}{2}(-D+n+1)+1 ;\\
		&\frac{\sqrt{(A_3 t_1+A_5)^2-4 A_2(A_1 t_1^2+A_4 t_1+A_6)}-2 A_2 t_n-A_3 t_1-A_5}{2 \sqrt{(A_3 t_1+A_5)^2-4 A_2(A_1 t_1^2+A_4 t_1+A_6)}}\Bigg)\\
		+
		& \frac{(2 A_1 x(2 A_ 2 t_n+A_5)+A_4(2A_2 t_n-A_3 x+A_5)-A_3 t_n(A_3 t_1+A_5)-2 A_3 A_6)}{(A_3 t_1+A_5)^2-4 A_2(A_1 t_1^2+A_4 t_1+A_6)} \\
		& \left(A_1 t_1^2 + A_2 t_n^2 + A_3t_1t_n + A_4 t_1 + A_5 t_n + A_6\right)^{\frac{1}{2}(-D+n+1)-1}.
	\end{aligned}
	\label{eq:proof 03}
\end{equation}
Substituting \eqref{Gt-nn}, we have
\begin{equation}
	\begin{aligned}
		&\sum_{r\neq n}\frac{\partial}{\partial t_r} \mathcal{G}_{\WH{n},\WH{n}}(\boldsymbol{t})=\sum_{r\neq n}\frac{\partial}{\partial t_r}\left( {F^*_{nn}(\bm{t})\over F^*_{nn}(\bm{t}=0)}\right)^{{(D-n)\over 2}}\\
		 =&\frac{D-n}{2}F^*_{nn}(\boldsymbol{t})^{\frac{D-n}{2}-1}\left((F^*_{nn}(\boldsymbol{t}=0))^{\frac{n-D}{2}}\sum_{r\neq n}\frac{\partial }{\partial t_r} F^*_{nn}(\boldsymbol{t})\right).
	\end{aligned}
\end{equation}
From \textbf{Lemma 01} and \textbf{Lemma 03} of Appendix \ref{sec:prop of F}, one can see $\left((F^*_{nn}(\boldsymbol{t}=0))^{\frac{n-D}{2}}\sum_{r\neq n}\frac{\partial }{\partial t_r} F^*_{nn}(\boldsymbol{t})\right)$ is a constant independent of $\boldsymbol{t}$. Thus, substituting \eqref{eq:proof 02} and \eqref{eq:proof 03},
with some computations \eqref{eq:proof 01} becomes
\begin{equation}
	\begin{aligned}
0 =& H_1\cdot { }_2 F_1\Bigg(\frac{1}{2}(D-n-1)+1, \frac{1}{2}(-D+n+1) ; \frac{1}{2}(-D+n+1)+1 ;\\
		&\frac{\sqrt{(A_3 t_1+A_5)^2-4 A_2(A_1 t_1^2+A_4 t_1+A_6)}-2 A_2 t_n-A_3 t_1-A_5}{2 \sqrt{(A_3 t_1+A_5)^2-4 A_2(A_1 t_1^2+A_4 t_1+A_6)}}\Bigg)+H_0,
	\end{aligned}
\end{equation}
where
\bea		H_0& =&\Bigg(\frac{(2 A_1 x(2 A_ 2 t_n+A_5)+A_4(2A_2 t_n-A_3 x+A_5)-A_3 t_n(A_3 t_1+A_5)-2 A_3 A_6)}{(A_3 t_1+A_5)^2-4 A_2(A_1 t_1^2+A_4 t_1+A_6)}F^*_{nn}(\boldsymbol{t})-F^*_{1n}(\boldsymbol{t})\Bigg)\nn
		&\times & |F|(\boldsymbol{t})^{-\frac{D-1-n}{2}-1},\eea
and
\bea		H_1&=&\frac{D-n}{1-D+n}|F|(\boldsymbol{t})^{-\frac{D-1-n}{2}}\frac{1}{\sqrt{(A_3 t_1+A_5)^2-4 A_2(A_1 t_1^2+A_4 t_1+A_6)}}\nn
		&& \times  \left(\frac{\sqrt{(A_3 t_1+A_5)^2-4 A_2(A_1 t_1^2+A_4 t_1+A_6)}+2 A_2 t_n+A_3 t_1+A_5}{2 \sqrt{(A_3 t_1+A_5)^2-4 A_2(A_1 t_1^2+A_4 t_1+A_6)}}\right)^{\frac{1}{2}(D-n-1)}\nn
	& & 	\times \left(\frac{\partial F^*_{nn}(\boldsymbol{t})}{\partial t_1}+F^*_{nn}(\boldsymbol{t})\frac{8A_1A_2t_1+4A_2A_4-2A_3^2t_1-2A_3A_5}{(A_3 t_1+A_5)^2-4 A_2(A_1 t_1^2+A_4 t_1+A_6)}\right)
	\eea
Next, we only need to prove that $H_1=H_0=0$. One can check
\begin{equation}
	\begin{aligned}
		&A_2=\frac{1}{2}\frac{\partial |F|(\boldsymbol{t})}{\partial t_n^2},~~~~
		A_3t_1+A_5=\frac{\partial^2|F|(\boldsymbol{t})}{\partial t_n}-2 A_2 \cdot t_n,\\
		&A_1 t_1^2+A_4 t_1+A_6=|Q|-\frac{\partial|F|(\boldsymbol{t})}{\partial t_n} \cdot t_n+A_2 t_n^2
	\end{aligned}
\end{equation}
Thus we have
\bea
		&&(A_3 t_1+A_5)^2-4 A_2(A_1 t_1^2+A_4 t_1+A_6)
		=\left(\frac{\partial|F|(\boldsymbol{t})}{\partial t_n}-2 A t_n\right)^2-4 A_2\left(|F|(\boldsymbol{t})-\frac{\partial|F|(\boldsymbol{t})}{\partial t_n} t_n+A_2 t_n^2\right) \nn
		&=&\frac{\partial|F|(\boldsymbol{t})}{\partial t_n} \frac{\partial|F|(\boldsymbol{t})}{\partial t_n}-4 A_2|F|(\boldsymbol{t})
		=4\left(\sum_{i,j=1}^n F_{i n}^* \cdot F_{j n}^*+\sum_{i, j=1}^{n-1} F_{i n, j n}^*|Q|\right)
		=4\left(\sum_{i, j=1}^n F_{i j}^*\right) F_{n n}^*,
	\label{eq:proof 04}
\eea
where the last line of the above equation is obtained from \textbf{Lemma 04} of Appendix \ref{sec:prop of F}. Furthermore, we have
\bea
	&	&(2 A_1 x(2 A_ 2 t_n+A_5)+A_4(2A_2 t_n-A_3 x+A_5)-A_3 t_n(A_3 t_1+A_5)-2 A_3 A_6)\nn
	&	=&\left(A_5+A_3 t_1+2 A_2 t_n\right)\left(A_4+2 A_1 t_1+A_3 t_n\right)-2 A_3|F|(\boldsymbol{t})\nn
	&	=&\frac{\partial|F|(\boldsymbol{t})}{\partial t_n} \frac{\partial|F|(\boldsymbol{t})}{\partial t_1}-2 \frac{\partial}{\partial t_1}\left(\frac{\partial}{\partial t_n}|F|(\boldsymbol{t})\right) \cdot|Q|=4\left(\sum_{i, j=1}^n F_{1 i}^* F_{j n}^*-\sum_{j \neq 1,i \neq n} F^*_{1 j,  in }|F|(\boldsymbol{t})\right)\nn &=& 4 F_{1 n}^* \cdot \sum_{i, j=1}^{n} F_{i j}^*
	\label{eq:proof 05}
\eea
and
\bea	&	&8A_1A_2t_1+4A_2A_4-2A_3^2t_1-2A_3A_5=
		-\frac{\partial}{\partial t_1}\left((A_3 t_1+A_5)^2-4 A_2(A_1 t_1^2+A_4 t_1+A_6)\right)\nn
		&=&-4\left(\sum_{i, j=1}^n F_{i j}^*\right) \frac{\partial F_{n n}^*}{\partial t_1}.
	\label{eq:proof 06}
\eea
{ The last line of \eqref{eq:proof 06} is obtained by that $\sum_{i, j=1}^n F_{i j}^*$ is a constant due to \textbf{Lemma 03} of Appendix \ref{sec:prop of F}}. By \eqref{eq:proof 04} \eqref{eq:proof 05} and \eqref{eq:proof 06}, one can easily check that $H_1=H_0=0$. Thus  we have completed the proof.


\section{The decomposition of scalar integrals}\label{sca-red}

In this part, we will show that for $d$-dimensional spacetime, the one-loop scalar integrals with  more than $(d+1)$  propagators can be decomposed 
as the linear combination of scalar integrals with $(d+1)$ propagators. This
problem has been addressed in \cite{sca-red} and we repeat here for reader's convenience. To make
the discussion more concrete, we will focus on the case $d=4$.  

With the translation invariance of loop momentum, we can write the scalar
integral with $(n+1)$ propagators as 
\be
I_{n+1}\equiv \int {d^D\ell\over i\pi^{D/2}}{1\over \prod_{i=0}^n ((\ell-K_i)^2-M_i^2)}~,~K_0=0,~~~~~\label{sr-1-1}
\ee
where $D=4-2\eps$ for the dimensional regularization. There are $n$ external momenta $K_i$ in \eref{sr-1-1} living in $d$-dimensional spacetime. For $d=4$, 
when $n\geq 5$, we can take any five momenta, for example $(K_1, K_2, K_3, K_4, K_5)$, and write down 
following  linear relation for them:
\bea
0& =& \eps(K_1, K_2, K_3,K_4)K_5+\eps(K_2, K_3, K_4,K_5)K_1+\eps(K_3, K_4, K_5,K_1)K_2\nn & & +\eps(K_4, K_5, K_1,K_2)K_3+\eps(K_5, K_1,K_2, K_3)K_4~~\label{sr-1-2}
\eea
where $\eps(K_1, K_2, K_3,K_4)=\epsilon_{\mu_1\mu_2\mu_3\mu_4}K_1^{\mu_1}K_2^{\mu_2}K_3^{\mu_3}K_4^{\mu_4}$. This relation can be easily proved by proper contraction with total antisymmetric tensor $\epsilon_{\mu_1\mu_2\mu_3\mu_4}$. For example, contracting with $\epsilon_{\rho_1\rho_3\rho_5\mu}K_1^{\rho_1}K_3^{\rho_3}K_5^{\rho_5}$ at the both sides of \eref{sr-1-2}, the right hand side gives
\be \eps(K_3, K_4, K_5,K_1)\eps(K_1, K_3, K_5, K_2)+\eps(K_5, K_1,K_2, K_3)\eps(K_1, K_3, K_5, K_4)~~\label{sr-1-3}\ee
which is zero by the total antisymmetry. 

Now we consider the following combination
\bea & & \sum_{i=0}^5 \a_i ((\ell-K_i)^2-M_i^2)=(\sum_{i=0}^5 \a_i)\ell^2
-2\ell\cdot (\sum_{i=1}^5 \a_i K_i)+ \sum_{i=0}^5 \a_i (K_i^2-M_i^2).~~\label{sr-1-4}\eea
By taking $\a_i$ for $i=1,...,5$ to be the coefficients given in the relation \eref{sr-1-2} and $\a_0=-\sum_{i=1}^5 \a_i$, \eref{sr-1-4} becomes
\bea & &  \sum_{i=0}^5 \a_i (K_i^2-M_i^2)~~\label{sr-1-5}\eea
Using this result, we can see that 
\bea
I_{n+1} &= &  \int {d^D\ell\over i\pi^{D/2}}{1\over \prod_{i=0}^n ((\ell-K_i)^2-M_i^2)} \times {\sum_{i=0}^5 \a_i ((\ell-K_i)^2-M_i^2)\over \sum_{i=0}^5 \a_i (K_i^2-M_i^2)}\nn
& = & {1\over \sum_{i=0}^5 \a_i (K_i^2-M_i^2)}\sum_{j=0}^5
 \int {d^D\ell\over i\pi^{D/2}}{\a_j\over \prod_{i=0,i\neq j}^n ((\ell-K_i)^2-M_i^2)}~~~~~\label{sr-1-6}
\eea
Thus we have decomposed the scalar integral $I_{n+1}$ as the linear combination
of six scalar integrals $I_{n}$. We can iteratively decompose $I_{n}$ further by exactly the same method until all scalar integrals are pentagons. 

Noticing that the above decomposition is purely algebraic, thus even with the generating function defined in \eref{gen-1-1}, one can reduce the case with arbitrary $n$ to the case with $n\leq d+1$, which is the condition  stated in the main body.

\section{Solving $\mathcal{H}_{n+1 ; j}$ with Mathematica}\label{H-Mathematica}

\subsection{Functions of Solving $\mathcal{H}_{n+1 ; j}$}
The Mathematica notebook "Hmatrix.nb" is structured to solve the matrix $\mathcal{H}_{n+1 ; j}$ in section \ref{H-matrix}. This involves defining the propagators, specifying the external momenta, constructing the Cayley matrix representation, and calculating the determinants of the required submatrices.

The function $\mathcal{H}_{n+1 ; j}$ is implemented in Mathematica as \texttt{HH[n1, i, r, c]}. It gives the reduction coefficient of the integral indicated by the column index $c$ to the basis indicated by the row index $r$. More explicitly, the four arguments are:

\begin{itemize}
    \item \textbf{n1}: The number of external legs ({ $n1 = n + 1$} for the $(n+1)$-gon). \newline
    \textit{Range}: Positive integers (typically { 1 to 5} for up to 5-point functions in four dimension). \newline
    \textit{Example}: For a box diagram, {$n=3,\; n1 = 4$}.

    \item \textbf{i}: The index of the propagator whose power is two. \newline
    \textit{Range}: $0$ to {$n$}. \newline
    \textit{Meaning}: Indicates which propagator's power is raised from 1 to 2. \newline
    \textit{Example}: $i = 2$ means the third propagator's power is increased to 2.

    \item \textbf{r}: The list of removed propagators for the row of the $H$ matrix. \newline
    \textit{Range}: Any subset of \{0, 1, \dots, {$n$}\}. \newline
    \textit{Meaning}: Specifies which propagators are removed in the denominator. \newline
    \textit{Example}: \{0, 2\} means the first and third propagators are removed.

    \item \textbf{c}: The list of removed propagators for the column of the $H$ matrix. \newline
    \textit{Range}: Any subset of \{0, 1, \dots, {$n$}\}. \newline
    \textit{Meaning}: Specifies which propagators are removed in the denominator. \newline
    \textit{Example}: \{1\} means the second propagator is removed.
\end{itemize}

The modified Cayley matrix $F$ is defined with elements:

\begin{equation}
 f_{ij} = (K_i - K_j)^2 - M_i^2 - M_j^2, \quad i, j = 0, \dots, n
\end{equation}

$\Delta_I$ represents the determinant of the sub-matrix obtained from the modified Cayley matrix $F$ by extracting the rows and columns corresponding to the index set $I$.

\subsection{Example of function \texttt{HH[n1, i, r, c]}}

To use the "Hmatrix.nb" notebook effectively, please ensure that you meet the following requirements:
\begin{itemize}
\item \textbf{Software}: Mathematica 12.0 or above is required. The notebook utilizes functions and features that are only available in these or newer versions.
\item \textbf{Notebook}: Download \texttt{Hmatrix.nb} and place it in a desired directory.
\end{itemize}

\textbf{Dependencies:} No additional external packages are required. All computations are performed using built-in Mathematica functionalities.

\begin{tcolorbox}[colback=white, colframe=black!75, title=Example 1: Tadpole (1-point function)]
  \textbf{Input:} \texttt{HH[1, 0, \{\}, \{\}]}\\
  \textbf{Output:}
  \begin{equation*}
  \frac{2 - D}{\Delta_{\text{tad}}}
  \end{equation*}
\end{tcolorbox}  

{This output represents the element $\mathcal{H}_{1 ; 0}$, where $D$ denotes the spacetime dimension, $\Delta_{tad} = -2 M_0^2 $ is the determinant of the Cayley matrix $F_{tad}$ for this configuration.
}

\begin{tcolorbox}[colback=white, colframe=black!75, title=Example 2: Bubble (2-point function)]
  \textbf{Input:} \texttt{HH[2, 1, \{0\}, \{0\}]}\\
  \textbf{Output:}
  \begin{equation*}
    \frac{2-D}{f_{11}}
  \end{equation*}
  
\end{tcolorbox}

{This output corresponds to the element $\mathcal{H}_{2 ; 1}$, where the propagator labeled $\{0\}$ is removed from both the row and column in the matrix.}

\begin{tcolorbox}[colback=white, colframe=black!75, title=Example 3: Triangle (3-point function)]
  \textbf{Input:} \texttt{HH[3, 2, \{0, 1\}, \{0\}]}\\
  \textbf{Output:}
  \begin{equation*}
    -\frac{(2-D) f_{12}}{f_{22} \Delta_{23}}
  \end{equation*}
\end{tcolorbox}
{This output represents the element $\mathcal{H}_{3 ; 2}$, with the propagators labeled $\{0,1\}$removed from the rows and $\{0\}$ from the column.}

\begin{tcolorbox}[colback=white, colframe=black!75, title=Example 4: Box (4-point function)]
\textbf{Input:} \texttt{HH[4, 3, \{0, 1, 2\}, \{0\}]}\\
\textbf{Output:}
\begin{equation*}
  \frac{\left(f_{12} f_{13}-f_{11} f_{23}\right)\left(\frac{2-D}{\Delta_{24}}-\frac{(2-D) f_{13}}{f_{33} \triangle_{24}}\right)}{\Delta_{234}}+\frac{\left(-f_{13} f_{22}+f_{12} f_{23}\right)\left(\frac{2-D}{\Delta_{34}}-\frac{(2-D) f_{23}}{f_{33} \triangle_{34}}\right)}{\Delta_{234}}
\end{equation*}
\end{tcolorbox}

\subsection{Examples of Numerical or Analytic Value}
For numerical or partially symbolic results, the function \texttt{NumvalueCayleyElements} can be used to generate numerical substitutions for the relevant Cayley elements.

\begin{tcolorbox}[colback=white, colframe=black!75, title=Example of numerical value \texttt{HH[3, 0, \{1, 2\}, \{\}]}]
  \textbf{Input:}\\
                  \texttt{result5 = HH[3, 0, \{1, 2\}, \{\}];}\\
                  \texttt{inputData1 $=\left\{K_0 \rightarrow\{0,0,0,0\}, \dots, M_1 \rightarrow 1, M_2 \rightarrow 1\right\}$;}\\
                  \texttt{NumvalueCayleyElements [inputData1]}\\
  \textbf{Output:}
  \begin{equation*}
    \left\{f_{0 0} \rightarrow-8, f_{01} \rightarrow-5, f_{02} \rightarrow-7, f_{01} \rightarrow-5, \dots, \Delta_{13} \rightarrow-33, \Delta_{23} \rightarrow 0, \Delta_{t r i} \rightarrow 8\right\}
  \end{equation*}
  \textbf{Input:} \texttt{result5 /. NumvalueCayleyElements [inputData1]}\\
  \textbf{Output:} 
    \begin{equation*}
      \frac{1}{2}\left(-\frac{5}{72}(2-D)+\frac{2-D}{9}\right)+\frac{1}{2}\left(\frac{7(2-D)}{264}+\frac{1}{33}(-2+D)\right)
    \end{equation*}
\end{tcolorbox}

{Here, \texttt{result5} represents a specific configuration for a three-point function, set up with the arguments \texttt{HH[3, 0, \{1, 2\}, \{\}]}. The \texttt{inputData1} assigns values such as $K_0=\{0,0,0,0\}$, and $M_1=1,M_2=1$ for the masses. \texttt{NumvalueCayleyElements[inputData1]} computes the numerical values for elements in the Cayley matrix based on \texttt{inputData1}. After substituting the numerical values from \texttt{inputData1} into \texttt{result5}, we obtain a simplified expression in terms of $D$, evaluated with the given numerical parameters.}

For the analytic evaluation, the function \texttt{AnalyticCayleyElements} is used:

\begin{tcolorbox}[colback=white, colframe=black!75, title=Example of \texttt{AnalyticCayleyElements}]
  \textbf{Input:}\texttt{AnalyticCayleyElements[3]}\\
  \textbf{Output:} 
  \begin{equation*}
    \{f_{00}\to -2 M_0^2,f_{01}\to -M_0^2-M_1^2+\text{MP}\left(K_0,K_0\right)-2 \text{MP}\left(K_0,K_1\right)+\text{MP}\left(K_1,K_1\right),\dots\}\\
  \end{equation*}
  \textbf{Input:} \texttt{result5/.AnalyticCayleyElements[3]}\\
  \textbf{Output:} 
  The formula is too lengthy, please refer to "Hmatrix.nb" for the complete expression.  
\end{tcolorbox}
This progress computes the symbolic values of the Cayley matrix elements and determinants for the 3-point function. The $MP(K_i, K_j)$ term represents the Lorentz inner product with signature $\{1, -1, \dots, -1\}$.

\end{document}